\documentclass[11pt,english]{article}
\usepackage{jheppub}

\usepackage{amsmath}
\usepackage{amssymb}
\usepackage{amsfonts}

\usepackage{graphicx}

\usepackage{braket}
\usepackage{subfigure}
\usepackage{wrapfig}

\newcommand{\beq}{\begin{equation}}
\newcommand{\eeq}{\end{equation}}

\newcommand{\fft}[2]{\frac{#1}{#2}}
\newcommand{\nn}{{\nonumber}}
\newcommand{\m}{\mu}
\newcommand{\emm}{s}

\begin{document}

\preprint{MCTP-13-24}

\title{What do non-relativistic CFTs tell us about Lifshitz spacetimes?}

\author{Cynthia Keeler,}
\author{Gino Knodel}
\author{and James T. Liu}

\affiliation{Michigan Center for Theoretical Physics, Randall Laboratory of Physics,\\
The University of Michigan, Ann Arbor, MI 48109--1040, USA}

\emailAdd{keelerc@umich.edu}
\emailAdd{gknodel@umich.edu}
\emailAdd{jimliu@umich.edu}

\abstract{We study the reconstructability of $(d+2)$-dimensional bulk spacetime from $(d+1)$-dimensional boundary data, particularly concentrating on backgrounds which break $(d+1)$-dimensional Lorentz invariance.  For a large class of such spacetimes, there exist null geodesics which do not reach the boundary.  Therefore classically we expect some information is trapped in the bulk and thus invisible at the boundary.  We show that this classical intuition correctly predicts the quantum situation: whenever there are null geodesics which do not reach the boundary, there are also ``trapped scalar modes'' whose boundary imprint is exponentially suppressed. We use these modes to show that no smearing function exists for pure Lifshitz spacetime, nor for any flow which includes a Lifshitz region.  Indeed, for any (planar) spacetime which breaks $(d+1)$-dimensional Lorentz invariance at any radius, we show that local boundary data cannot reconstruct complete local bulk data.}

\maketitle

\section{Introduction}

For the past several years, there has been much interest in applying
the powerful field theory/gravity dualities developed in the late
90s and early 2000s to field theories without Lorentz invariance.
These non-relativistic forms of AdS/CFT, often collectively referred
to as AdS/CMT due to their relevance for condensed matter systems,
have provided a new tool for examining strongly coupled non-relativistic
systems (see, e.g.~\cite{Hartnoll:2009sz,McGreevy:2009xe,Huijse:2011hp,Sachdev:2010ch}
and references therein).

Although many aspects of the bulk-boundary dictionary familiar from
AdS/CFT carry forward to these systems without alteration, some aspects
differ strongly. The first obvious difference is the symmetry group;
since the goal is to consider spacetime duals to nonrelativistic systems,
the asymptotic symmetries of the spacetime should be nonrelativistic.
As a consequence, the spatial and temporal components of the metric
near the boundary must scale differently with the radius. This different
scaling in fact means the notion of a boundary itself is altered;
using the Penrose definition of a conformal boundary leads to a degenerate
boundary metric. However, more careful treatments have shown that
there is still a reasonable notion of a boundary as the location where
metric components go to infinity, and holographic calculations can
be performed using suitable prescriptions \cite{Baggio:2011cp,Papadimitriou:2010as,Papadimitriou:2011qb,Ross:2009ar,Ross:2011gu,Mann:2011hg,Chemissany:2012du}.

The spacetimes studied as possible nonrelativistic duals fall into
two main classes: those which have Lifshitz scaling symmetry, and
those which have the larger Schr\"odinger symmetry.  There are also
other spacetimes in the literature, including the warped AdS spacetimes
\cite{Anninos:2008fx,Chakrabarti:2009ww,Liu:2009kc,Anninos:2010pm,Compere:2013bya}, which exhibit temporo-spatial anisometry. In
this paper, we will concentrate on spacetimes which have Lifshitz
symmetry at least in some region, but many of our conclusions apply
to more general spacetimes with scaling differences between space
and time.

One of the best studied examples of a boundary-bulk duality system
with space/time anisotropy is the so-called Lifshitz spacetime, given
by
\begin{equation}
ds_{d+2}^{2}=-\left(\fft{L}r\right)^{2z}dt^{2}+\left(\fft{L}r\right)^{2}(d\vec{x}_d^{2}+dr^{2}).
\end{equation}
It was first proposed in \cite{Kachru:2008yh} and has been extensively
studied since. In order to remove some of
the concerns about degenerate boundary behavior, \cite{Braviner:2011kz,Singh:2010cj,Singh:2013iba}
have considered replacing the near-boundary UV region of the spacetime
with an asymptotically AdS spacetime. Other numerical constructions of these backgrounds are available in \cite{Goldstein:2009cv,Harrison:2012vy,Bhattacharya:2012zu,Knodel:2013fua}. Additionally, there are a set of ``hyperscaling-violating'' solutions which still have a Lifshitz-like symmetry, proposed in \cite{Narayan:2012hk} and studied further in \cite{Dong:2012se,Bueno:2012sd,Dey:2013oba,Edalati:2012tc}. We will consider an ansatz which allows for analysis of all these cases.

Much recent progress has been made in creating a complete bulk/boundary
dictionary for nonrelativistic systems \cite{Ross:2009ar,Baggio:2011cp,Ross:2011gu,Mann:2011hg}. In the well studied case of
Lorentzian AdS/CFT, an important part of this dictionary is the correspondence
between normalizable modes, which scale as $r^{\Delta_{+}}$ near
the boundary, and states in the Hilbert space of the dual field theory.
In particular, a quantized bulk field $\phi$ can be mapped to its
corresponding boundary operator $O$ via
\begin{equation}
\phi\mapsto O=\lim_{r\rightarrow0}r^{-\Delta_{+}}\phi\label{eq:bulk to boundary map}.
\end{equation}
 The remarkable fact here is that both operators can be quantized
in terms of the same creation/annihilation operators, which implies
an isomorphism between the Fock space representations of bulk and
boundary Hilbert spaces \cite{Balasubramanian:1998sn,Balasubramanian:1998de}.
Moreover, the map \eqref{eq:bulk to boundary map} can be inverted
in position space. As a result, local quantum fields in the bulk can
be expressed in terms of boundary operators with the help of a
so-called smearing function $K$ \cite{Bena:1999jv,Hamilton:2005ju,Hamilton:2006az}. Consequently, we can study CFTs to learn something about their gravitational duals
\cite{Balasubramanian:1999ri,Banks:1998dd,Fitzpatrick:2011jn}.

If AdS/CMT is to be understood as a `true' equivalence between a field
theory and a gravitational theory, rather then just a set of prescriptions
to compute condensed matter quantities, one should expect that a similar
statement can be made for nonrelativistic systems. In other words,
the field theory should somehow contain all the relevant information
about the gravitational theory. In this paper, we address this issue
by investigating the extent of the reconstructability of bulk information
from boundary data in nonrelativistic spacetimes.

\begin{figure}[t]
\centering
\includegraphics[height=6cm]{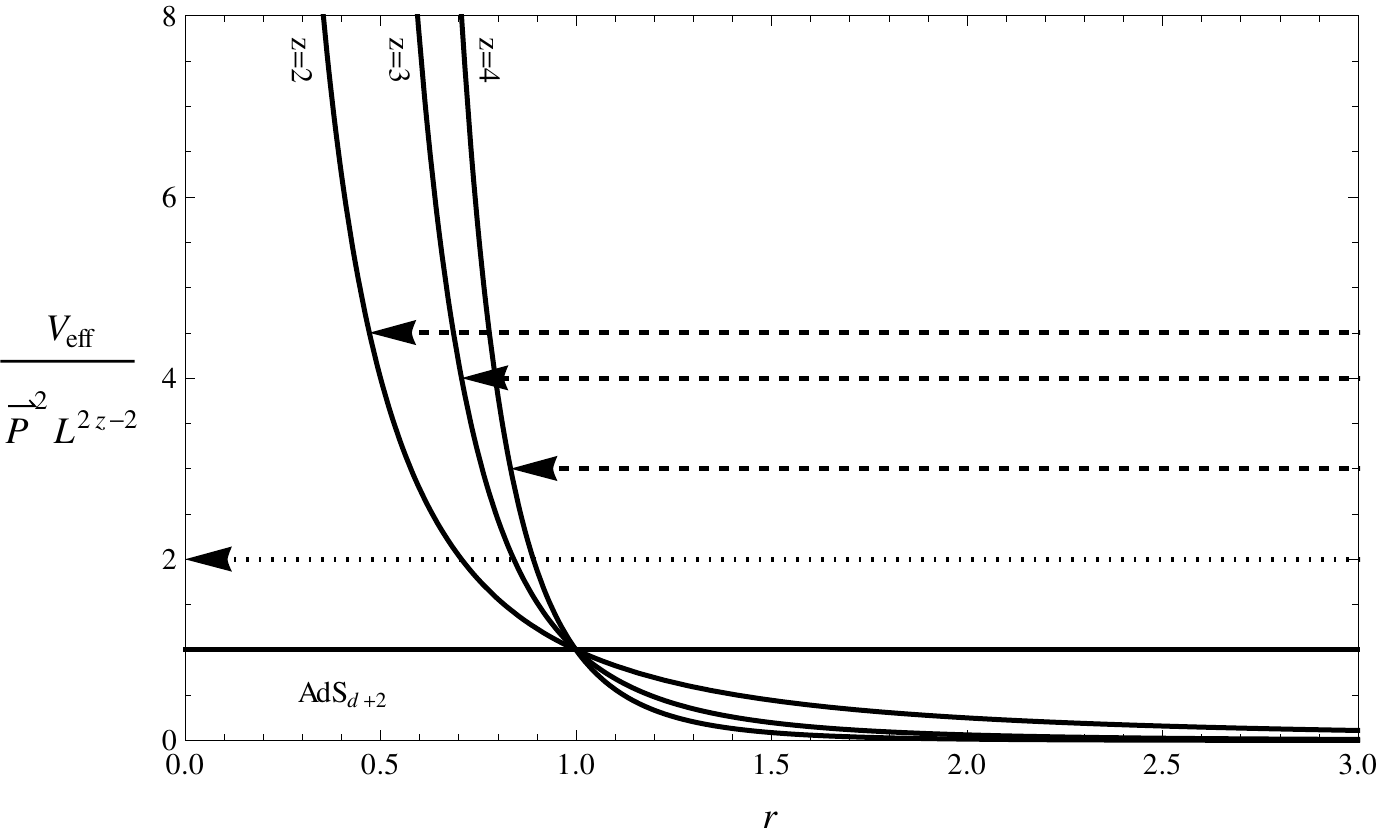}
\caption{\label{fig:Vpnull}Effective potential \eqref{eq:Veff geo for lif}
for null geodesics ($\kappa=0$) in AdS ($z=1$) and Lifshitz spacetimes
($z=2,3,4$). In Lifshitz, light rays sent from the bulk in any nonradial
direction have to turn around at finite $r$ and can never reach the
boundary.}
\end{figure}

A simple argument why this procedure is not straightforward can be made by
studying geodesics in the corresponding backgrounds. For Lifshitz
spacetime, the effective potential is given by
\begin{equation}
V_{{\rm eff}}(r)=\left(\fft{L}r\right)^{2z}\kappa+\left(\fft{L}r\right)^{2(z-1)}\vec{p}\,^{2}.
\label{eq:Veff geo for lif}
\end{equation}
Null geodesics ($\kappa=0$) with nonzero transverse momentum $p$ turn around at
finite $r$ and never reach the boundary (see Figures~\ref{fig:Vpnull}
and \ref{fig:nullcurves}).
This is a result of the nonrelativistic nature of the dual theory,
which manifests itself in the fact that the effective speed of light  $g_{tt}/g_{xx}$ diverges as $r\rightarrow0$. Therefore, information about the
transverse direction of the bulk geometry can never reach an observer
at the boundary.

Quantum mechanically the picture is different. In general,
wavefunctions are allowed to tunnel through any classically forbidden
region to reach the boundary, so there is hope that bulk reconstruction
is possible after all. However, as we will demonstrate, at large momenta
the imprint these tunneling modes leave at the boundary is exponentially
small and as a consequence, a smearing function cannot be constructed.
Our arguments closely follow those of \cite{Bousso:2012mh,Leichenauer:2013kaa},
where first steps towards generalizing smearing functions to spaces
other than pure AdS were made.

Our analysis for the case of pure Lifshitz spacetime can be easily
generalized to show that smearing functions do not exist for any geometry
that allows for `trapped modes', that is, modes that have to tunnel
through a momentum-barrier in the potential to reach the boundary. In \cite{Leichenauer:2013kaa}, the authors show that the smearing function in their spherically symmetric spacetimes can indeed become well-defined, at least in some bulk region, once they change from an AdS-Schwarzschild solution to a nonsingular asymptotically AdS spacetime.  Our case, however, does not allow such a resolution. Importantly, the smearing function in Lifshitz remains ill-defined everywhere if we resolve the tidal singularity \cite{Kachru:2008yh,Copsey:2010ya,Horowitz:2011gh}  into an $\mathrm{AdS}_{2}\times\mathbb{R}^{d}$ or $\mathrm{AdS}_{d+2}$ region.  It also remains ill-defined everywhere if we replace the near-boundary region with an asymptotic $\mathrm{AdS}_{d+2}$ region, or if we do both replacements at once \cite{Harrison:2012vy,Bhattacharya:2012zu,Knodel:2013fua}.

The problem we encounter when trying to construct a smearing function
is related to modes with large transverse momentum. Introducing a momentum-cutoff $\Lambda$, however, will force us to give up
the ability of reconstructing full bulk locality in the transverse
direction.

The outline of this paper is as follows: In section \ref{sec:classical},
we discuss the idea of bulk reconstruction via classical geodesics
in Lifshitz spacetimes. We show that there are null geodesics that
cannot reach the boundary. We generalize this statement to flows
involving Lifshitz regions, as well as more general nonrelativistic
spacetimes with planar symmetry, considering the constraints arising from the null energy condition. In section
\ref{sec:quantum}, we turn to the quantum picture and study solutions
of the scalar field equations for the same class of spacetimes. In
particular, we show analytically that for $z=2$ Lifshitz, there are
modes that have to tunnel through a momentum-barrier in the potential
to reach the boundary and are thus exponentially suppressed. We generalize
this result to arbitrary $z$ using the WKB approximation. In section
\ref{sec:sflifshitz}, we review the construction of smearing functions
via the mode-sum approach and attempt to construct a Lifshitz smearing
function. Using WKB methods, we show that this attempt fails due to
the existence of `trapped modes', which have exponentially small boundary
imprint. In section \ref{sec:sfgeneral}, we generalize our findings
to show that smearing functions do not exist for a large class of
nonrelativistic spacetimes. Finally, in section \ref{sec:modifyingbb}
we interpret our results and their implications for bulk locality.
We argue that only a hard momentum cutoff allows bulk reconstruction,
at the cost of giving up locality in the transverse direction.

\section{\label{sec:classical}The Classical Picture: Bulk reconstruction
via light signals }

We now set our notation and discuss the classical paths of geodesics within the spacetimes we study.  Specifically, we consider planar metrics of the form
\begin{equation}
ds_{d+2}^{2}=-e^{2A(r)}dt^{2}+e^{2B(r)}d\vec{x}_{d}^{2}+e^{2C(r)}dr^{2}.\label{eq:metans}
\end{equation}
This ansatz is sufficiently general to include AdS, Lifshitz with
general $z$ (with or without hyperscaling violation), $\mathrm{AdS}_{2}\times\mathbb{R}^{d}$
and spacetimes which interpolate among them. Note that one of the
three functions $A,B$ and $C$ can always be eliminated by a suitable
gauge choice. However, it is convenient to keep these functions arbitrary
for now, so that we can more easily accommodate the various gauge
choices that have been used for AdS and Lifshitz metrics in the literature.
The metric (\ref{eq:metans}) can be trivially rewritten as
\begin{equation}
ds_{d+2}^{2}=e^{2B(r)}[-e^{2W(r)}dt^{2}+d\vec{x}_{d}^{2}]+e^{2C(r)}dr^{2}.\label{eq:metans with W,B,C}
\end{equation}
where we defined $W\equiv A-B$. For $W=0$, the $(d+1)$-dimensional
metric at constant $r$ is Lorentz invariant. This encompasses the
pure AdS case as well as Lorentz invariant domain wall flows. The
$W\neq0$ case allows for `non-relativistic' backgrounds such as pure
or asymptotic Lifshitz backgrounds as well as for planar black holes.
In this case, we may interpret $e^{-W}$ as the gravitational redshift
factor%
\footnote{Note that this assumes that there is an asymptotic reference region
where $W=0$, so that $(d+1)$-dimensional Lorentz invariance is restored.
This would occur, for example, in an AdS to Lifshitz flow.}.

\begin{figure}[t]
\begin{minipage}{.45\textwidth}
\centering
\includegraphics[height=6cm]{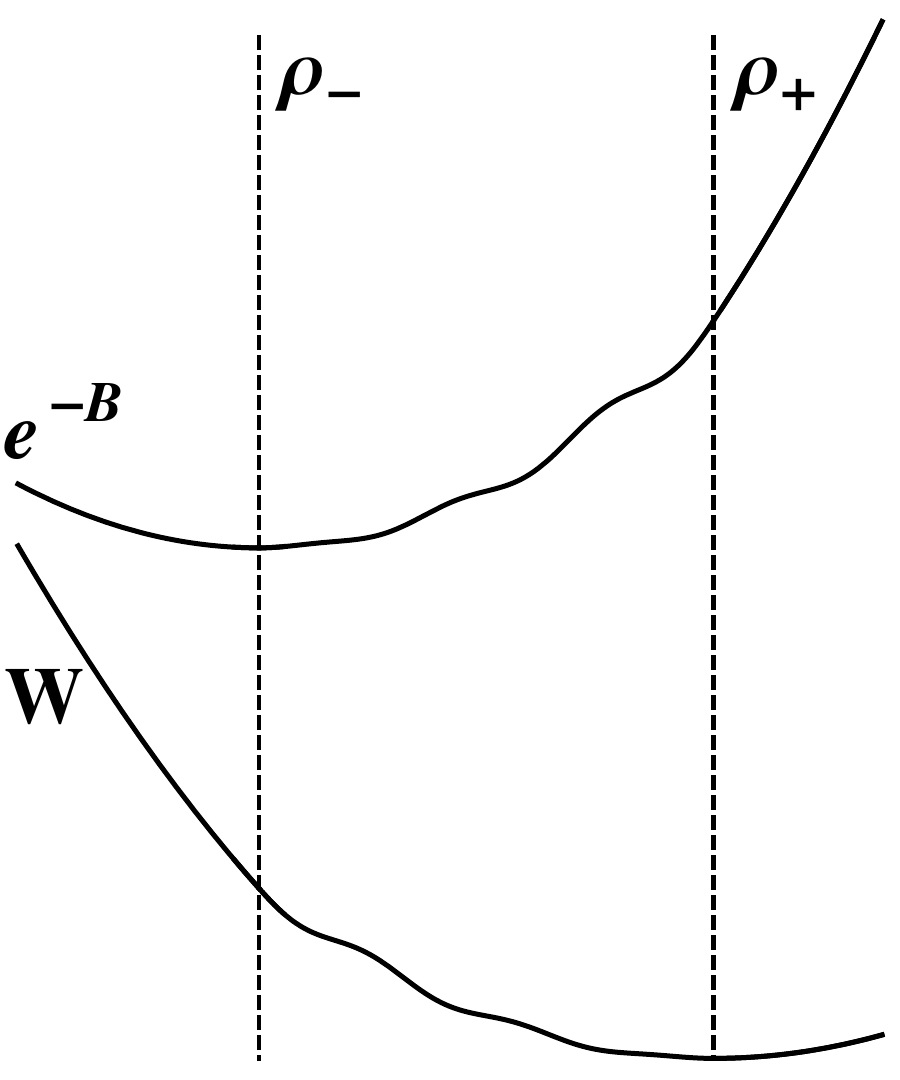}
\end{minipage}%
\begin{minipage}{.55\textwidth}
\centering
\includegraphics[height=5cm]{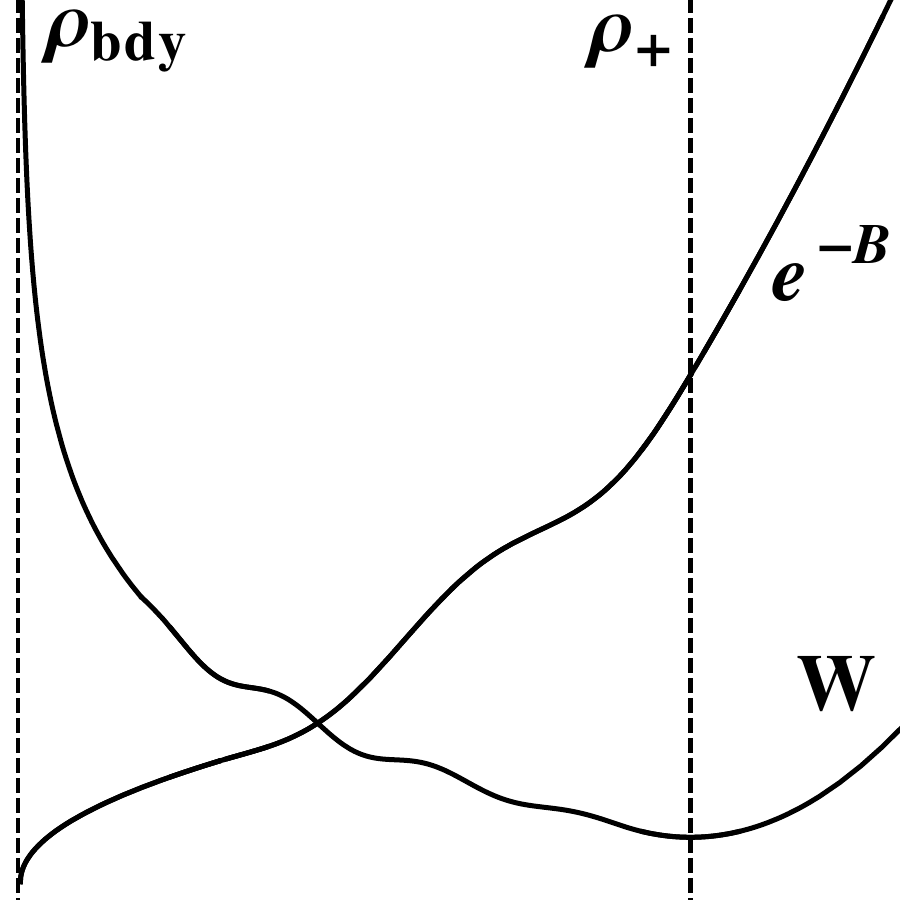}
\end{minipage}
\caption{\label{fig:possibleNEC}Two sketches of functions $W$ and $e^{-B}$ which
obey the null energy conditions (\ref{eq:nullrrsimple}) and (\ref{eq:nullxxsimple}). The figure on the right approaches Lifshitz asymptotics at $\rho_{\text{bdy}}$.}
\end{figure}

The global behavior of the metric is constrained by the null energy condition
(subsequently NEC; for previous work see \cite{Hoyos:2010at,Liu:2012wf}).
The two independent conditions are
\begin{align}
-R_{\, t}^{t}+R_{\, r}^{r} & =de^{W-C}\partial_{r}\left(-e^{-W-C}\partial_{r}B\right)\geq0,\label{eq:nullcondrr}
\\
-R_{\, t}^{t}+R_{\, x_{1}}^{x_{1}} & =e^{-W-(d+1)B-C}\partial_{r}\left(e^{W+(d+1)B-C}\partial_{r}W\right)\geq0.\label{eq:nullcondxx}
\end{align}
 Here $x_{1}$ is any one of the $\vec{x}$ transverse directions.
If we choose a gauge where $A=C$, or equivalently $W=C-B$, these
conditions simplify to
\begin{align}
\left((e^{-B})'e^{-2W}\right)' & \geq0,\label{eq:nullrrsimple}\\
\left(W'e^{dB}\right)' & \geq0,\label{eq:nullxxsimple}
\end{align}
where $'$ denotes derivatives with respect to the radial coordinate
$\rho$ in the corresponding gauge. Since $e^{dB}\geq0$, we can use
the second condition to deduce the following statements about $W$
(see Figure~\ref{fig:possibleNEC}):
\begin{align}
\text{If }W'|_{\rho_{-}} \leq 0 &\quad\Rightarrow\quad W'|_{\rho\leq\rho_{-}}\leq0;\nn\\
\text{If }W'|_{\rho_{+}}\geq 0 & \quad\Rightarrow\quad W'|_{\rho\geq\rho_{+}}\geq 0.
\end{align}
From (\ref{eq:nullrrsimple}) we can deduce similar equations for $e^{-B}$.
If we combine the two constraints (\ref{eq:nullrrsimple}) and (\ref{eq:nullxxsimple}), we learn
about the second derivatives
of $W$ and $e^{-B}$ when their first derivatives have the same sign:
\begin{align}
\text{If }W'|_{\rho_{-}} \leq 0\text{ and }(e^{-B})'|_{\rho_{-}} \leq 0, &\quad \Rightarrow\quad
W''|_{\rho\leq\rho_{-}} \geq 0\text{ and }(e^{-B})''|_{\rho\leq\rho_{-}} \geq 0;\nn\\
\text{If }W'|_{\rho_{+}} \geq 0\text{ and }(e^{-B})'|_{\rho_{+}} \geq 0, &\quad \Rightarrow\quad
W''|_{\rho\geq\rho_{+}} \geq 0\text{ and }(e^{-B})''|_{\rho\geq\rho_{+}} \geq 0.
\label{eq:concavity}
\end{align}
These conditions will constrain the bulk geometry, and in particular the behavior of the
redshift factor $e^{-W}$.

As mentioned in the introduction, we may gain insight about the bulk spacetime by considering
null geodesics.  Such geodesics are easily obtained by noting that the metric (\ref{eq:metans})
admits Killing vectors
\beq\label{eq:killings}
\fft{\partial}{\partial t},\qquad{}\fft{\partial}{\partial x^{i}}.
\eeq
This allows us to define the conserved energy and momentum
\begin{equation}
E\equiv e^{2A}\dot{t},\qquad\vec{p}\equiv e^{2B}\dot{\vec{x}},
\end{equation}
where a dot indicates a derivative with respect to the affine parameter
$\lambda$. Geodesics then obey
\begin{equation}
-\kappa=\left(\fft{ds}{d\lambda}\right)^{2}=-e^{-2(W+B)}E^{2}+e^{-2B}\vec{p}\,^{2}+e^{2C}\dot{r}^{2}.
\end{equation}
 If we define
\begin{equation}
V_{{\rm eff}}\equiv e^{2(W+B)}\kappa+e^{2W}\vec{p}^{2},
\label{eq:veff}
\end{equation}
with $\kappa=1$ for timelike and $\kappa=0$ for null geodesics, then we find
\begin{equation}
e^{2(W+B+C)}\dot{r}^{2}=E^{2}-V_{{\rm eff}}.
\end{equation}
This is of the form of an energy conservation equation, $E_{{\rm tot}}=E_{{\rm kin}}+V_{{\rm eff}}$,
where
\begin{equation}
E_{\mathrm{kin}}=e^{2(W+B+C)}\dot{r}^{2}.
\end{equation}
\subsection{\label{sub:lifgeos}Lifshitz geodesics}

We now study specifically Lifshitz spacetimes.  Pure Lifshitz spacetime corresponds to taking
\begin{equation}
W=-(z-1)\log(r/L),\qquad B=-\log(r/L),\qquad C=-\log(r/L)
\end{equation}
in the metric ansatz \eqref{eq:metans with W,B,C}. Note that the
`horizon' is at $r=\infty$, while the boundary is at $r=0$. The
effective potential for geodesics is
\begin{equation}
V_{{\rm eff}}(r)=\left(\fft{L}r\right)^{2z}\kappa+\left(\fft{L}r\right)^{2(z-1)}\vec{p}\,^{2}.
\label{eq:Veff lifshitz}
\end{equation}
The behavior of the second term depends on the value of $z$. For
$z=1$, this term is a constant, and just shifts the overall potential.
For $z>1$, the second term still grows as $r^{-2(z-1)}$, but this
growth is slower than that of the $\kappa$ term. In addition, it
vanishes at the horizon, $r\rightarrow\infty$. For null geodesics
($\kappa=0$), the effective potential is completely determined by
this term.

\begin{figure}[t]
\centering
\subfigure[Looking from a future vantage point.]{ \includegraphics[width=0.45\textwidth]{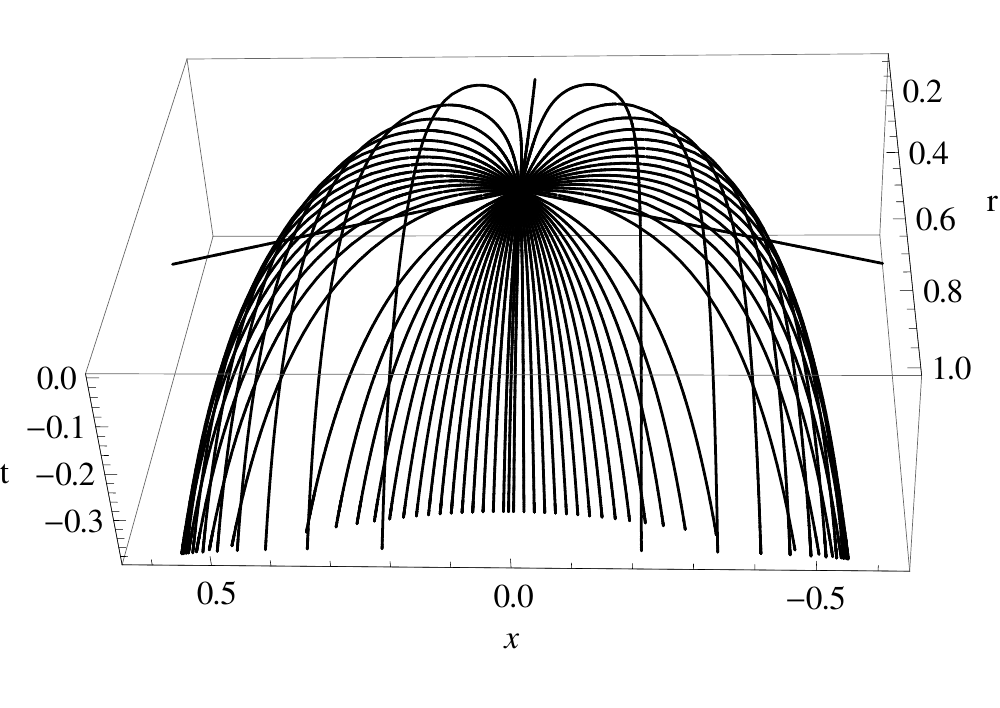}}
\subfigure[A ``side'' view.]{ \includegraphics[width=0.45\textwidth]{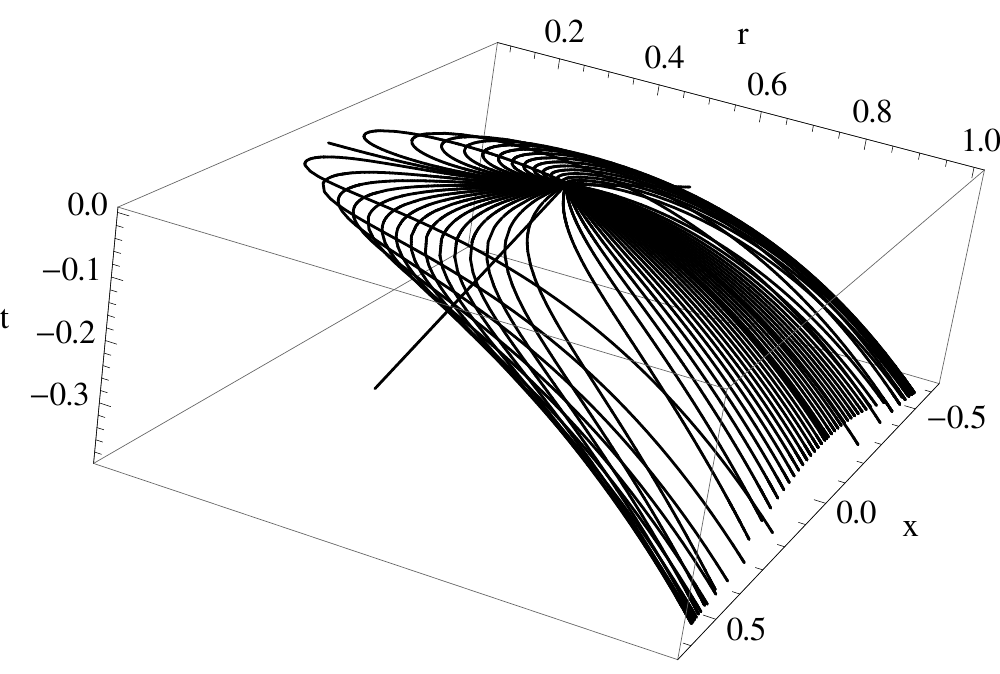}}
\caption{\label{fig:nullcurves}Plot of null curves through the point $t=0$,
$\vec{x}=0$, $r=1/2$ in Lifshitz space with $z=3$.}
\label{fig:Lifgeos}
\end{figure}

Radial null geodesics ($\vec{p}=0$) do not feel any effective
potential. For $z>1$, non-radial geodesics on the other hand cannot
reach the boundary. In Figure~\ref{fig:Lifgeos}, we have plotted
several such light rays which all converge on one point in space;
these rays delineate the causal past of that point. As we can see
in the figure, only the null geodesic which stays at constant $x=0$
can reach the boundary at $r=0$; all others turn around at some minimum
$r$.

The result is that a full classical%
\footnote{Note that `classical' in this case refers to the geometric optics
limit, as opposed to just the $N\rightarrow\infty$ limit.}
reconstruction of the bulk from the boundary is not possible. An
observer at the boundary can never receive any signals from the bulk which travel
with a nonzero momentum in the transverse direction. Consequently,
this observer will not be able to `resolve' transverse length scales in the bulk.
Of course, this picture is somewhat naive and cannot be taken as
a proof that bulk reconstruction is impossible. However, as we
will show in the next section, the picture carries forward to the
quantum case, even though tunneling through classically forbidden regions
is possible.

Two comments are in order at this point. First, notice that pure Lifshitz
spacetime has a pathology at $r\rightarrow\infty$. An infalling extended
object experiences infinitely strong tidal forces. To see this, consider
two parallel radial geodesics with energy $E$ travelling in the background
\eqref{eq:metans with W,B,C}. The geodesic deviation equation for
the transverse separation $X^{i}$ reads
\begin{equation}
\frac{D^{2}X^{i}}{Dt^{2}}=X^{i}E^{2}e^{-2\left(W+B+C\right)}\left[-B^{\prime}\left(W^{\prime}+C^{\prime}\right)+B^{\prime\prime}-\frac{\kappa}{E^{2}}e^{2\left(W+B\right)}\left(B^{\prime}\left(B^{\prime}-C^{\prime}\right)+B^{\prime\prime}\right)\right].
\end{equation}
 For Lifshitz spacetime, we have
\begin{equation}
\frac{D^{2}X^{i}}{Dt^{2}}=X^{i}\frac{E^{2}}{L^2}\left[\left(1-z\right)\left(\frac{r}{L}\right)^{2z}-\frac{\kappa}{E^{2}}\right].
\end{equation}
 For $z\neq1$, the relative acceleration diverges near the horizon
and the result is an infinitely strong tidal force. By now, there are
several known ways to resolve this issue \cite{Harrison:2012vy,Bhattacharya:2012zu,Knodel:2013fua,Bao:2012yt}.
For solutions which involve a running dilaton, a natural resolution
is to avoid the singularity by deforming the geometry such that it flows
to $\mathrm{AdS}_{2}\times\mathbb{R}^{d}$ in the deep infrared. More generally,
one can imagine several possible IR deformations that change the behavior of
the metric functions $W$, $B$ and $C$ at large $r$. These deformations
have to be consistent with the NECs in (\ref{eq:nullrrsimple}) and (\ref{eq:nullxxsimple}) above.
However, it is clear that while these procedures might cure
the problems encountered near the horizon, they do not change the
fact that geodesics sent towards the boundary still cannot overcome
the Lifshitz barrier \eqref{eq:Veff lifshitz}.

On the other hand, one could imagine that deforming the geometry in
the UV might help null geodesics to reach the boundary. Deformations which replace the UV with an AdS region have the benefit of clarifying the holographic prescription%
\footnote{See, however \cite{Baggio:2011cp,Papadimitriou:2010as,Papadimitriou:2011qb,Ross:2009ar,Ross:2011gu,Mann:2011hg}
for different approaches to holography in Lifshitz spacetimes.%
}.
If we imagine a geometry that is approximately
Lifshitz at some $\rho_{-}$, then $W'(\rho_-)<0$. The NECs thus dictate that $e^{W}$ has to
either continue increasing or asymptote to a constant as $\rho\rightarrow0$. The latter
case would correspond to an AdS to Lifshitz flow. For fixed transverse
momentum $p$, geodesics with large enough energy can now escape the
potential and reach the boundary. However, at fixed $E$, the height
of the potential barrier is controlled by $p^{2}$, so geodesics with
large transverse momentum remain trapped inside the bulk.

We conclude that for any spacetime that is approximately Lifshitz
in some region, part of the information about the bulk will always
be hidden from a classical boundary observer. The part that is missing
describes physics at large $p$, or equivalently small transverse
length scales. Again, we will see in the subsequent sections
that this statement has an exact equivalent in the quantum case.

\section{\label{sec:quantum}The Quantum Picture: Bulk reconstruction for
scalar fields }

While the geometric optics picture of the previous section already
captures some important physical properties of nonrelativistic gauge/gravity
dualities, a full analysis of the problem of bulk reconstruction from
the boundary clearly requires a treatment of quantum operators. To
this end, we consider solutions to the scalar field equations
and investigate what kind of imprint they can leave at the boundary. Specifically, we examine the amplitude of scalar modes near the UV boundary in terms of the size of fluctuations deep in the IR.

We begin by studying the Klein-Gordon equation for a scalar in the fixed background (\ref{eq:metans with W,B,C})
\begin{equation}
[e^{-W-(d+1)B-C}\partial_{M}e^{W+(d+1)B+C}g^{MN}\partial_{N}-m^{2}]\phi=0.\label{eq:KGe}
\end{equation}
Because of the Killing vectors (\ref{eq:killings}) present in our
metric ansatz, the wave equation is separable and we can write
\begin{equation}
\phi(t,\vec{x},r)=e^{i(\vec{p}\cdot\vec{x}-Et)}f(r).
\end{equation}
Then the Klein-Gordon equation (\ref{eq:KGe}) becomes
\begin{equation}
\left[e^{2(W+B-C)}\left(\partial_{r}^{2}+\fft{d(W+(d+1)B-C)}{dr}\partial_{r}\right)+E^{2}
-e^{2W}\vec{p}\,^{2}-e^{2(W+B)}m^{2}\right]f=0.
\end{equation}
Let us choose a gauge where $A=C$, or $W=C-B$. Equivalently, starting in
any given gauge we can introduce a new radial coordinate $\rho$ such
that
\begin{equation}
e^{C-B-W}dr=d\rho.\label{eq:dxdr}
\end{equation}
Note that $\rho$ is a tortoise coordinate for our metric ansatz.
This gives
\begin{equation}
[\partial_{\rho}^{2}+dB'\partial_{\rho}+E^{2}-e^{2W}\vec{p}\,^{2}-e^{2\left(W+B\right)}m^{2}]f=0,
\end{equation}
where primes denote derivatives with respect to $\rho$. If we now
let
\begin{equation}
f=e^{-dB/2}\psi,\label{eq:spsidef}
\end{equation}
 we end up with a Schr\"odinger-type equation
\begin{equation}
-\psi''+U\psi=E^{2}\psi,\label{eq:schr}
\end{equation}
where
\begin{equation}
U=V_{m}+V_{p}+V_{{\rm cos}},
\end{equation}
with
\begin{equation}
V_{m}= e^{2(W+B)}m^{2},\qquad
V_{p}= e^{2W}\vec{p}\,^{2},\qquad
V_{\mathrm{cos}}=(d/2)B''+(d/2)^{2}B'^{2}.
\end{equation}
Here $V_{m}$ and $V_{p}$ together form the effective potential \eqref{eq:veff} for geodesics,
with $\kappa$ replaced by $m^{2}$. The third term, $V_{\mathrm{cos}}$, is an additional
`cosmological' potential that is absent in the classical picture.

\subsection{\label{sub:scalarslifshitz}Scalars in Lifshitz spacetime}

For Lifshitz backgrounds, the Schr\"odinger potential can be written
as
\begin{equation}
U=\left(\fft{L}{z\rho}\right)^{2}\left(m^{2}+\fft{d(d+2z)}{4L^{2}}\right)
+\left(\fft{L}{z\rho}\right)^{2(1-1/z)}\vec{p}\,^{2},
\end{equation}
where we introduced a new radial coordinate according to \eqref{eq:dxdr}.
Explicitly, we have
\begin{equation}
\rho=\fft{L}z\left(\fft{r}L\right)^{z}.\label{eq:rho from r lifshitz}
\end{equation}
Note that both $V_{m}$ and the entirety of $V_{{\rm cos}}$ contribute
to the $1/\rho^{2}$ blowup as $\rho\to0$ (corresponding to the boundary).
The fact that these two pieces scale with the same power of $\rho$
is a feature of Lifshitz spacetime; it will not continue to be true
for more complicated spacetimes such as the AdS-Lifshitz flows studied
in section \ref{sub:adslifads}.

The qualitative behavior of solutions to the Schr\"odinger equation
is roughly as follows: The wavefunction starts out oscillating deep
in the bulk ($\rho\rightarrow\infty$) and crosses the potential barrier
at the classical turning point $\rho_{0}$. For $\rho<\rho_{0}$, the mode must tunnel
under the barrier, and thus
the wavefunction will in general be a superposition of exponentially
growing and suppressed modes. We will only be interested in the mass
ranges where the growing solution is non-normalizable. Thus, the normalizable
modes relevant for canonical quantization are exponentially suppressed in the area of this barrier
at small $\rho$.

For $z=1$, $V_{p}$ is a constant, but for $z>1$ it blows up near
the boundary, although less fast than the other terms in the potential.  Specifically,
$V_p/V_m \propto e^{-2B}$.  For spacetimes with Lifshitz asymptotics,
\beq
\partial_\rho \left(e^{-B}\right)\biggr|_{\rho_{\text bdy}}=\left.\partial_\rho\left(\frac{z\rho}{L}\right)^{1/z}\right|_{\rho_{\text bdy}}>0.
\eeq
Consequently, $\partial_\rho e^{-B}>0$ throughout the spacetime.  Near the boundary, the mass term $V_m$ will always dominate, but $V_p$ will increase in relative importance as we head in towards the IR region.

Because of the different behavior
of the mass/cosmological and momentum-dependent terms, it is crucial to distinguish between two qualitatively different
`types' of tunneling. If at a given energy, the momentum $\vec{p}$
is sufficiently small, the wavefunction crosses the barrier at a point
where $V_{p}$ is subdominant compared to the other terms in the potential.
Consequently, the $1/\rho^{2}$ part of $U$ will control
the suppression near the boundary. We shall refer to those modes as
\textit{free modes}. This name is justified, because even though they
are tunneling, classically they correspond to null geodesics that
can reach the boundary.

If $\vec{p}$ is large,
the wavefunction crosses the barrier already at a point where $U\approx V_{p}$,
and the wavefunction will receive an additional suppression by an
exponential in $\vec{p}$, due to tunneling through this thicker barrier. We shall refer to this class of solutions
as \textit{trapped modes}. They play a crucial role in our analysis,
as they are the quantum equivalent to nonradial null-geodesics that
cannot reach the boundary.

We may study the behavior of these free and trapped modes by solving the Schr\"odinger
equation \eqref{eq:schr} in a Lifshitz background.  It is convenient to scale out the energy
$E$ by introducing the dimensionless coordinate
\begin{equation}
\zeta=E\rho.\label{eq:zetadef}
\end{equation}
Then \eqref{eq:schr} becomes $-\psi''(\zeta)+(U-1)\psi(\zeta)=0$
where
\begin{equation}
U=\fft{\nu_{z}^{2}-1/4}{\zeta^{2}}+\fft\alpha{\zeta^{k}},\label{eq:dimpot}
\end{equation}
with
\begin{equation}
\nu_{z}=\fft1z\sqrt{(mL)^{2}+(d+z)^{2}/4},\qquad\alpha=\left(\fft{EL}z\right)^{k}\biggl(\fft{\vec{p}}E\biggr)^{2},\qquad k=2(1-1/z).\label{eq:nualphadef}
\end{equation}
Since the null energy condition demands $z\ge1$, we generally focus
on the case $0<k<2$. (The $k=0$, or pure AdS, case is familiar and
can be treated by standard methods.) In this case, the boundary ($\zeta\to0$)
behavior of $U$ is $\sim1/\zeta^{2}$, while the horizon ($\zeta\to\infty$)
behavior is $\sim1/\zeta^{k}$.

Near the boundary, we have
\begin{equation}
-\psi''+\fft{\nu^{2}-1/4}{\zeta^{2}}\psi\approx0\qquad\Rightarrow\qquad\psi\sim A\zeta^{1/2-\nu}+B\zeta^{1/2+\nu}.\label{eq:ABdef}
\end{equation}
 Using \eqref{eq:rho from r lifshitz}, \eqref{eq:zetadef} and (\ref{eq:spsidef}),
we can express the behavior of the original Klein-Gordon
field in terms of the original coordinate $r$ as
\begin{equation}
\phi\sim\hat{A}\left(\fft{r}L\right)^{\Delta_{-}}+\hat{B}\left(\fft{r}L\right)^{\Delta_{+}},
\end{equation}
where
\begin{equation}
\hat{A}=A\left(\fft{EL}z\right)^{1/2-\nu},\quad\hat{B}=B\left(\fft{EL}z\right)^{1/2+\nu},\quad\Delta_{\pm}=\fft{d+z}2\pm\sqrt{(mL)^{2}+\left(\fft{d+z}2\right)^{2}}.
\end{equation}

We will consider only the mass range where the first solution (related to $A$) is non-normalizable
with respect to the Klein-Gordon norm, while the second solution (related to $B$) is
normalizable. Via the AdS/CFT correspondence, non-normalizable modes
represent classical sources of an operator $O$ at the boundary, which
redefine the Hamiltonian of the field theory \cite{Maldacena:1997re,Gubser:1998bc,Witten:1998qj}.
Normalizable fluctuations are placed on top of these classical sources
and they correspond to different states in the field theory, or equivalently
expectation values of $O$ \cite{Balasubramanian:1998de,Balasubramanian:1998sn}.%
\footnote{Lifshitz spacetimes present some subtleties when considering alternate quantizations.  The range of masses for which both boundary conditions are normalizable is larger than in the AdS case, but modes which would not be normalizable in AdS (but apparently are in Lifshitz) suffer from a novel instability. Particularly in these cases it appears more difficult to redefine the Hamiltonian in the usual way \cite{Keeler:2012mb,Andrade:2012xy,Andrade:2013wsa}.}
We will only be interested in the situation where the
boundary Hamiltonian is fixed, so we will consequently treat non-normalizable
solutions as non-fluctuating.  The fluctuating modes to be quantized are thus the
normalizable modes given by $B$.  As a result, we will end up setting $A=0$ and
investigating the consequences of doing so%
\footnote{Note that this is in contrast with the computation of AdS/CFT correlators, where
$B$ is interpreted as the response to turning on a source $A$.}.

Turning now to the horizon, we see that both terms in \eqref{eq:dimpot} fall off as
$\zeta\to\infty$.  Hence the horizon behavior is given by%
\footnote{For simplicity, we have assumed $1<k<2$.  For $0<k\le1$, the horizon falloff
$\sim1/\zeta^k$ is insufficiently fast, and the potential becomes long-ranged.  This introduces
a correction to the horizon behavior of the wavefunction.  However, this is unimportant for our
discussion, as we have no need for the asymptotic phase of $\psi$ in the classically
allowed region.}
\begin{equation}
-\psi''-\psi\approx0\qquad\Rightarrow\qquad\psi\sim ae^{i\zeta}+be^{-i\zeta}.\label{eq:sabdef}
\end{equation}
In terms of the original $r$ coordinate, this becomes
\begin{equation}
\psi\sim a\exp\left(i\fft{EL}z\left(\fft{r}L\right)^{z}\right)+b\exp\left(-i\fft{EL}z\left(\fft{r}L\right)^{z}\right),
\end{equation}
so that
\begin{equation}
\phi\sim a\left(\fft{r}L\right)^{d/2}\exp\left(i\fft{EL}z\left(\fft{r}L\right)^{z}\right)+b\left(\fft{r}L\right)^{d/2}\exp\left(-i\fft{EL}z\left(\fft{r}L\right)^{z}\right).
\end{equation}
The horizon modes correspond to infalling and outgoing waves, given by $a$ and $b$, respectively.
Since the wave equation is second order and linear, the boundary data $(A,B)$ must be linearly
related to the horizon data $(a,b)$.  AdS/CFT correlators are generally computed by taking infalling
conditions at the horizon, corresponding to $b=0$, while bulk normalizable modes
are given instead by taking $A=0$ at the boundary.  Of course, the precise relation between
boundary and horizon data can only be obtained by solving the wave equation.  While
this cannot be performed in general, the exact solution is known for $z=2$, where the potential
$U$ is analytic.  We now turn to this case, as it provides a clean example of the behavior of
trapped modes and in particular the exponential suppression that they receive when
tunneling under the barrier in the potential.

\subsection{A Specific Example: $z=2$ Lifshitz}
\label{subsec:z=2Lif}

For a pure Lifshitz background with $z=2$, or $k=1$, the potential
(\ref{eq:dimpot}) is analytic in $\zeta$ and the Schr\"odinger equation
takes the form
\begin{equation}
-\psi''+\left(\fft{\nu^{2}-1/4}{\zeta^{2}}+\fft\alpha\zeta-1\right)\psi=0,
\end{equation}
where $\alpha=\vec p\,^2L/2E$.
As this is essentially Whittaker's equation, the solution can be written in terms of
the Whittaker functions $M_{-i\alpha/2,\nu}(-2i\zeta)$ and $W_{-i\alpha/2,\nu}(-2i\zeta)$,
or equivalently in terms of confluent hypergeometric functions \cite{Kachru:2008yh}.
Expanding for $\zeta\to0$ and demanding that $\psi$ satisfies the boundary asymptotics
\eqref{eq:ABdef} for normalizable and nonnormalizable modes gives
\begin{align}
\psi =&
\left[
\left(\fft{i}2\right)^{\fft12+\nu}B-\left(\fft{i}2\right)^{\fft12-\nu}\fft{\Gamma(-2\nu)
\Gamma(\fft12+\nu+\fft{i\alpha}2)}{\Gamma(2\nu)\Gamma(\fft12-\nu+\fft{i\alpha}2)}A\right]
M_{-i\alpha/2,\nu}(-2i\zeta)\nn\\
&+\left[
\left(\fft{i}2\right)^{\fft12-\nu}\fft{\Gamma(\fft12+\nu+\fft{i\alpha}2)}{\Gamma(2\nu)}A
\right]W_{-i\alpha/2,\nu}(-2i\zeta).
\label{eq:MWpsi}
\end{align}
For the horizon, we expand for large $\zeta$ and compare with \eqref{eq:sabdef} to
obtain
\begin{eqnarray}
\psi&=&\left[e^{-\pi\alpha/4}\fft{\Gamma(\fft12+\nu+\fft{i\alpha}2)}{\Gamma(1+2\nu)}2^{-i\alpha/2}
b\right]M_{-i\alpha/2,\nu}(-2i\zeta)\nn\\
&&+\left[e^{\pi\alpha/4}2^{i\alpha/2}a+e^{i\pi(\fft12-\nu)}
e^{\pi\alpha/4}\fft{\Gamma(\fft12+\nu+\fft{i\alpha}2)}{\Gamma(\fft12+\nu-\fft{i\alpha}2)}
2^{-i\alpha/2}b\right]W_{-i\alpha/2,\nu}(-2i\zeta).
\label{eq:MWhoriz}
\end{eqnarray}
Comparing \eqref{eq:MWpsi} with \eqref{eq:MWhoriz} gives the relation between horizon
and boundary coefficients
\begin{eqnarray}
A & = & (2i)^{\fft12-\nu}\fft{\Gamma(2\nu)}{\Gamma(\fft12+\nu-\fft{i\alpha}2)}e^{\pi\alpha/4}\left(2^{-i\alpha/2}b-e^{i\pi(\fft12+\nu)}\fft{\Gamma(\fft12+\nu-\fft{i\alpha}2)}{\Gamma(\fft12+\nu+\fft{i\alpha}2)}2^{i\alpha/2}a\right),\nn
\\
B & = & (2i)^{\fft12+\nu}\fft{\Gamma(-2\nu)}{\Gamma(\fft12-\nu-\fft{i\alpha}2)}e^{\pi\alpha/4}\left(2^{-i\alpha/2}b-e^{i\pi(\fft12-\nu)}\fft{\Gamma(\fft12-\nu-\fft{i\alpha}2)}{\Gamma(\fft12-\nu+\fft{i\alpha}2)}2^{i\alpha/2}a\right).\label{eq:ABfromab}
\end{eqnarray}
Although we are primarily interested in normalizable modes in the Lifshitz bulk, we first note
that the usual computation of the retarded Green's function proceeds by taking infalling
boundary conditions at the horizon, namely $b=0$.  Then \eqref{eq:MWhoriz} immediately
gives
\begin{equation}
\psi_{\rm infalling}\sim W_{-i\alpha/2,\nu}(-2i\zeta).
\end{equation}
We now demand that the coefficient of $M_{-i\alpha/2,\nu}(-2i\zeta)$ in \eqref{eq:MWpsi}
vanishes, from which we obtain
\begin{equation}
G_R(E,\vec p\,)\sim\fft{\hat B}{\hat A}=\left(\fft{EL}2\right)^{2\nu}\fft{B}A
=\left(\fft{EL}i\right)^{2\nu}\fft{\Gamma(-2\nu)
\Gamma(\fft12+\nu+\fft{i\alpha}2)}{\Gamma(2\nu)\Gamma(\fft12-\nu+\fft{i\alpha}2)},
\end{equation}
in agreement with \cite{Kachru:2008yh} when continued to Euclidean space.  Note that
in the large momentum limit, $p\to\infty$ (or more precisely for $\alpha\gg\nu$), the Whittaker
function $W_{-i\alpha/2,\nu}(-2i\zeta)$ is only large near the boundary, and decays
exponentially into the bulk.  This matches with the heuristic picture of AdS/CFT, where
the CFT `lives' on the boundary.  In the relativistic case, corresponding to an AdS geometry,
the boundary data has a power law falloff as it penetrates into the bulk.  However, for
this Lifshitz geometry, the falloff is exponential.

Of course, for the bulk reconstruction that we are interested in, we actually want to consider
the space of normalizable modes, as they are the ones that span the Hilbert space in the
bulk.  From the Hamiltonian picture, the natural norm is the Klein-Gordon norm, which is
in fact compatible with the norm for the Schr\"odinger equation \eqref{eq:schr}.
Normalizable modes correspond to taking $A=0$, so that
\begin{equation}
\psi_{\rm normalizable}\sim M_{-i\alpha/2,\nu}(-2i\zeta).
\end{equation}
Comparing \eqref{eq:MWpsi} with \eqref{eq:MWhoriz} then gives the relation between
bulk and boundary coefficients for normalizable modes
\begin{equation}
\fft{B}b=2^{-i\alpha/2}\left(\frac{2}{i}\right)^{\frac{1}{2}+\nu}\frac{\Gamma(\fft12+\nu+\fft{i\alpha}2)}{\Gamma\left(1+2\nu\right)}e^{-\pi\alpha/4}.
\label{eq:Bfrombexact}
\end{equation}
Note that $M_{-i\alpha/2,\nu}(-2i\zeta)$ is essentially a standing wave solution in the
classically allowed region $\zeta>\zeta_0$, where $\zeta_0$ is the classical turning point.
Since this interval is semi-infinite, the wavefunction must be normalized by fixing the
amplitude $b$ of these oscillations.  Hence the ratio $B/b$ is a direct measure of the
amplitude of properly normalized wavefunctions at the boundary.

Recall our previous distinction between the two different types of tunneling solutions: `free'
vs.\ `trapped' modes.
Modes with small momenta $p$ at fixed $E$ ($\alpha\ll\nu$) are `free modes'.  For these modes,
we have, up to an overall phase
\begin{equation}
\frac{|B|}{|b|}\approx\frac{2^{\nu+\frac{1}{2}}\Gamma\left(\frac{1}{2}+\nu\right)}
{\Gamma\left(1+2\nu\right)}.
\end{equation}
The tunneling process produces the typical scaling behavior $\sim\rho^{\Delta_{+}}$
at the boundary, but there is no exponential suppression. For large
momenta ($\alpha\gg\nu$) the modes are `trapped', and we find instead
\begin{equation}
\frac{|B|}{|b|}\approx\frac{\sqrt{4\pi}e^{-\left(\nu+\frac{1}{2}\right)}}
{\Gamma\left(1+2\nu\right)}\alpha^{\nu}e^{-\pi\alpha/2}.
\label{eq:B/b large alpha}
\end{equation}
These modes have to tunnel not only through the $1/\rho^{2}$
potential near the boundary, but also through the wider momentum barrier
$V_{p}\sim p^{2}/\rho$ at larger $\rho$. This causes the
solution to be exponentially suppressed when it reaches the boundary.
We conclude that the $z=2$ Lifshitz metric allows for `trapped modes',
which have arbitrarily small boundary imprint for large $p$.

Clearly, we could have obtained the exponential suppression factor $e^{-\pi\alpha/2}$
in \eqref{eq:B/b large alpha}
by simply setting $V_{m}=V_{\mathrm{cos}}=0$ in the Schr\"odinger
potential. More generally, since the size of $V_{p}$ is controlled
by $p^{2}$, in any interval {[}$\rho_{1}$,$\rho_{2}${]} away from
the boundary, i.e.\ in any region where the potential $U$ is bounded,
at large enough $p$ the difference in amplitudes between the points
$\rho_{1}$ and $\rho_{2}$ will always be governed by an exponential
relation like \eqref{eq:B/b large alpha}. For the purpose of determining
whether or not trapped modes exist in a given spacetime, it will therefore
be enough to study the equivalent tunneling problem in the potential
$U\equiv V_{p}$. We will come back to this issue later.

\subsection{WKB Approximation}

In order to study the existence of trapped modes in spacetimes beyond exact $z=2$ Lifshitz, it will be useful to have a formalism that provides
a qualitative description of the behavior of tunneling modes even
for cases where an analytic solution might not exist. This will allow us to study
Lifshitz with $z\neq2$, as well as more general backgrounds \eqref{eq:metans with W,B,C} with nontrivial $W$,
$B$ and $C$. The WKB method provides us with just such a formalism.
We make the standard ansatz
\begin{equation}
\psi\sim\frac{1}{\sqrt{P(\rho)}}e^{\int d\rho^{\prime}P(\rho^{\prime})}.
\end{equation}
For slowly-varying potentials, we can plug this back into \eqref{eq:schr}
and solve perturbatively for $P$. The details of this calculation
can be found in appendix~\ref{sec:WKB}. To lowest order, $P^{2}\approx U-E^{2}$
and the solution interpolates between an oscillating region in the
bulk and a tunneling region near the boundary. More explicitly, we
have
\begin{equation}
\psi\left(\zeta\right)=\begin{cases}
\left(U-E^{2}\right)^{-\frac{1}{4}}\left[Ce^{S(\rho)}+De^{-S(\rho)}\right], & \rho<\rho_{0};\\
\left(E^{2}-U\right)^{-\frac{1}{4}}\left[ae^{i\Phi(\rho)}+be^{-i\Phi(\rho)}\right], &\rho>\rho_{0},
\end{cases}\label{eq:WKB ansatz}
\end{equation}
where $\rho_{0}$ is the classical turning point and we defined the
action $S\left(\rho\right)=\int_{\rho}^{\rho_{0}}d\rho^{\prime}\sqrt{U-E^{2}}$
and a phase $\Phi\left(\rho\right)=\int_{\rho_{0}}^{\rho}d\rho^{\prime}\sqrt{E^{2}-U}$.
For potentials that behave as $U\sim 1/\rho^{2}$ near the
boundary (which includes both asymptotically AdS and Lifshitz spacetimes),
one has to include an additional correction term $U\rightarrow U+1/(2\rho)^{2}$
(See appendix~\ref{sec:WKB} for more details). Using the WKB
matching procedure between the two asymptotic regions, we find
\begin{eqnarray}
C & = & \left(e^{-i\frac{\pi}{4}}a+e^{i\frac{\pi}{4}}b\right),\nonumber
\\
D & = & \frac{i}{2}\left(e^{-i\frac{\pi}{4}}a-e^{i\frac{\pi}{4}}b\right).
\end{eqnarray}
The exponential growth/decay of the solution in the classically forbidden
region is manifest in the dependence on $S$ in \eqref{eq:WKB ansatz},
which roughly corresponds to the area of the tunneling barrier. The wider/higher
the barrier, the larger the corresponding factor $e^{S}$ is. We are
only interested in the normalizable, or decaying solution near the
boundary, so we will have to set $C=0$. Up to a finite error, the
WKB approximation then accurately captures the boundary behavior of
this solution, and in particular the exponential suppression between
bulk and boundary amplitudes%
\footnote{Notice however that calculating the ratio $B/A$,  which is needed
to calculate the standard field theory Green's function, would not
be possible. This is due to the fact that for a general solution,
the normalizable solution $\sim e^{-S}$ can `hide' under the non-normalizable
part $\sim e^{S}$, which grows much faster as $\rho\rightarrow0$.}.
%

\begin{figure}[t]
\centering
\includegraphics[width=0.6\textwidth]{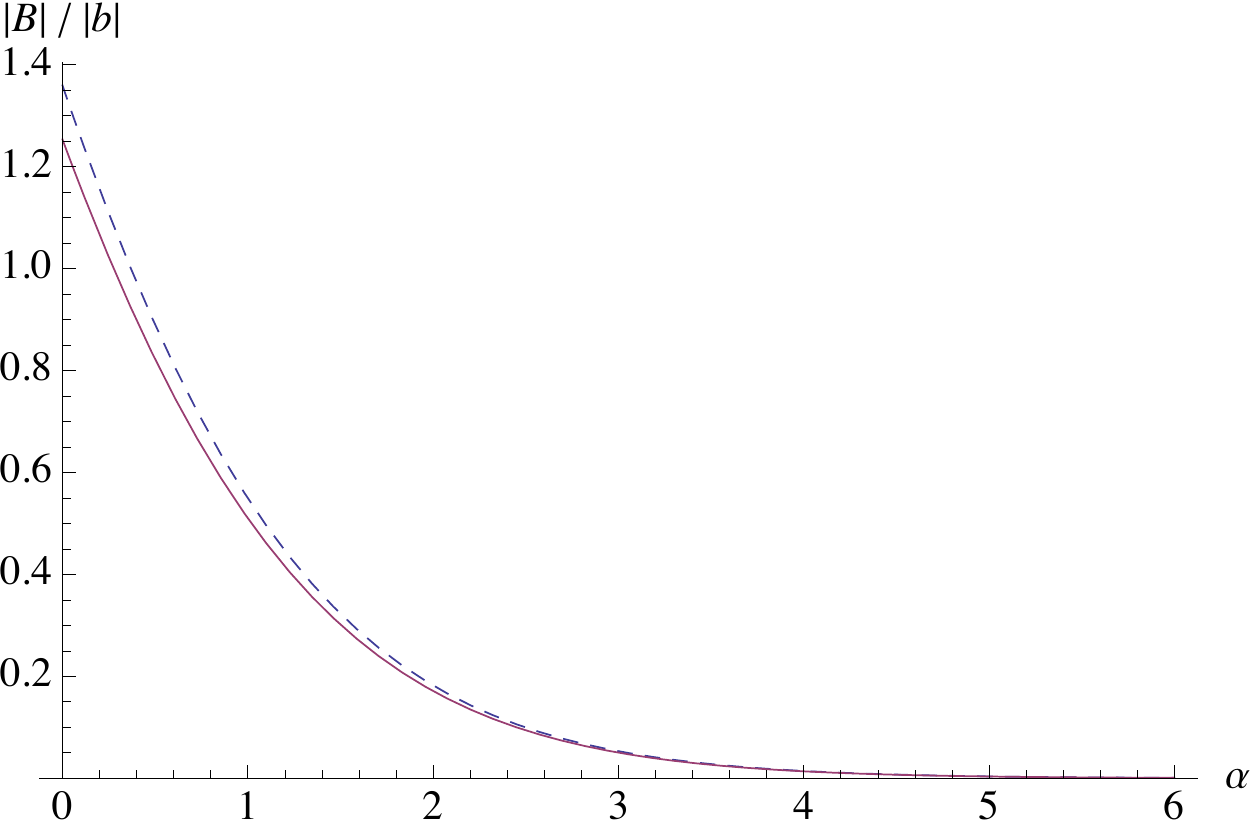}
\caption{\label{fig:nu=1z=2}
Plot of the WKB (dashed) and exact (solid) boundary normalization factor
$|B|/|b|$ as a function of $\alpha$.  Here we have taken $z=2$ and $\nu=1$.  The
large $\alpha$ behavior is exponentially suppressed, $|B|/|b|\sim \alpha^\nu e^{-\pi\alpha/2}$.}
\end{figure}

We can compare this WKB approximation with the exact solution for $z=2$ from section \ref{subsec:z=2Lif}. Figure~\ref{fig:nu=1z=2}
shows a plot of the WKB solution for $z=2$ Lifshitz, compared to the exact solution. As we can see,
the WKB approximation accurately captures the exponential momentum-suppression at large
$\alpha$.  (See also appendix~\ref{sec:WKB} for further `benchmark tests'.) In the
next section, we will use the WKB formalism to investigate for which
spacetimes smearing functions exist.

\section{\label{sec:sflifshitz}Smearing Functions in Lifshitz spacetimes }

In this section, we introduce smearing functions as a way to reconstruct
bulk physics from boundary dynamics. Using the WKB formalism developed
in appendix~\ref{sec:WKB}, we will show that for Lifshitz spacetimes, and more
generally for any flow involving Lifshitz, such reconstruction is
not possible.

First, recall that the normalizable solutions of the Klein Gordon
equation can be used to construct the Hilbert space of the bulk theory
in the following way: We decompose the scalar as
\begin{equation}
\phi\left(t,\vec{x},r\right)=\int dEd^{d}p\frac{1}{N_{E,p}}\left(\phi_{E,p}\left(t,\vec{x},r\right)a_{E,p}+\phi_{E,p}^{*}\left(t,\vec{x},r\right)a_{E,p}^{\dagger}\right),\label{eq:phi expansion}
\end{equation}
where $a_{E,p}$ are operators, $N_{E,p}\equiv\Braket{\phi_{E,p},\phi_{E,p}}^{\frac{1}{2}}$and
$\Braket{\cdot,\cdot}$ is the Klein-Gordon inner product, defined
by
\begin{equation}
\Braket{f,g}\equiv i\int_{\Sigma}d^{d}xdr\sqrt{-g}g^{00}\left(f^{*}\partial_{t}g-\left(\partial_{t}f^{*}\right)g\right).\label{eq:KGNorm}
\end{equation}
Here, the integral is to be taken over a spacelike slice $\Sigma$.%
\footnote{This norm accords with the norm preserved by the effective Schr\"odinger equation in (\ref{eq:schr}).
}%

If we choose $\Braket{\phi_{E,p},\phi_{E,p}^{*}}=0$, i.e.\ pick definite
frequency solutions, the $a$ and $a^{\dagger}$ are the usual creation/annihilation
operators for particles with wavefunction $\phi_{E,p}$. We can create
all possible states in the Fock space by repeatedly acting with $a^{\dagger}$
on the vacuum $\Ket{0}_{\mathrm{AdS}}$. In Lorentzian AdS/CFT, the
bulk-boundary dictionary states that there exists a boundary operator
defined by
\begin{equation}
O\left(t,\vec{x}\right)\equiv\lim_{r\rightarrow0}r^{-\Delta_{+}}\phi\left(t,\vec{x},r\right),\label{eq: O from phi}
\end{equation}
which is sourced by the classical, non-normalizable solution $\phi_{\mathrm{cl}}$
behaving as $r^{\Delta_{-}}$ at the boundary. Taking the
above limit in \eqref{eq:phi expansion}, we arrive at
\begin{equation}
O\left(t,\vec{x}\right)=\int dEd^{d}p\frac{1}{N_{E,p}}\left(\varphi_{E,p}\left(t,\vec{x}\right)a_{E,p}+\varphi_{E,p}^{*}\left(t,\vec{x}\right)a_{E,p}^{\dagger}\right).
\end{equation}
Here $\varphi_{E,p}\equiv\lim_{r\rightarrow0}r^{-\Delta_{+}}\phi_{E,p}$.
The remarkable fact is that the boundary operator can be expanded
in terms of \textit{the same} $a$,$a^{\dagger}$ as the bulk field.
Thus, to create an arbitrary state in the bulk we can use either bulk
operators or boundary operators that are `smeared' over $\vec{x}$
and $t$ in an appropriate way. For example, for a single-particle
state we have
\begin{equation}
a_{E,p}=\int dt^{\prime}d^{d}x^{\prime}N_{E,p}\varphi_{E,p}^{*}\left(t^{\prime},\vec{x}^{\prime}\right)O\left(t^{\prime},\vec{x}^{\prime}\right),\label{eq:a from O}
\end{equation}
 so the state $\Ket{E,p}_{\mathrm{AdS}}$ can be built entirely out
of boundary operators, and so on. Here we need to assume that the
$\varphi$ are normalized such that
\begin{equation}
\int dEd^{d}p\varphi_{E,p}^{*}\left(t,\vec{x}\right)\varphi_{E,p}\left(t^{\prime},x^{\prime}\right)=\delta\left(t-t^{\prime}\right)\delta\left(\vec{x}-\vec{x}^{\prime}\right).\label{eq:boundary normalization}
\end{equation}
Notice that \eqref{eq:boundary normalization}, and not \eqref{eq:KGNorm},
is the relevant inner product here. This is because the $\varphi_{E,p}$ are
not solutions to any equation of motion at the boundary; rather, they are
a set of complete functions%
\footnote{In other words: $O$ is an off-shell operator.}.
The condition \eqref{eq:boundary normalization} is not in tension
with the Klein-Gordon normalization condition in the bulk, since we
have explicitly factored out $N_{E,p}$ in \eqref{eq:phi expansion}.

Equation \eqref{eq:a from O} induces an isomorphism between the Fock-space
representations of the bulk and boundary Hilbert spaces. The question
we would like to answer is whether we can express any operator in
the bulk entirely in terms of boundary operators. In particular, we
would like to reconstruct $\phi$ from its corresponding boundary
operator $O$. We make the ansatz

\begin{equation}
\phi\left(t,\vec{x},r\right)=\int dt^{\prime}d^{d}x^{\prime}K\left(t,\vec{x},r|t^{\prime},\vec{x}^{\prime}\right)O\left(t^{\prime},\vec{x}^{\prime}\right),\label{eq:phi ansatz}
\end{equation}
where $K$ is called a smearing function. We can plug \eqref{eq:a from O}
back into \eqref{eq:phi expansion} to obtain:
\begin{equation}
K\left(t,\vec{x},r|t^{\prime},\vec{x}^{\prime}\right)=\int dEd^{d}p\phi_{E,p}\left(t,\vec{x},r\right)\varphi_{E,p}^{*}\left(t^{\prime},\vec{x}^{\prime}\right).\label{eq:candidate K}
\end{equation}
Note that this $K$ differs from the usual bulk-to-boundary propagator in that it is a relationship among normalizable modes.
Throughout this paper, we will assume that $K$ has a well-defined
Fourier transform, which allows us to interchange the order of integration
above. We will comment on some mathematical details and the precise
definition of $K$ in section \ref{sec:modifyingbb}.

In Lifshitz spacetime, the normalizable solutions are given by
\begin{equation}
\phi_{E,p}=e^{-i\left(Et-\vec{p}\cdot\vec{x}\right)}f_{E,p}=e^{-i\left(Et-\vec{p}\cdot\vec{x}\right)}e^{-\frac{d}{2}B}\psi_{E,p}.
\end{equation}
 Near the boundary,
\begin{equation}
\psi\approx B_{E,p}\zeta^{\frac{1}{2}+\nu}\equiv\hat{B}_{E,p}r^{z\left(\frac{1}{2}+\nu\right)},
\end{equation}
so that
\begin{equation}
\varphi_{E,p}=\lim_{r\rightarrow0}r^{-\Delta_{+}}\phi=e^{-i\left(Et-\vec{p}\cdot\vec{x}\right)}\hat{B}_{E,p}.
\end{equation}
 The normalization condition \eqref{eq:boundary normalization} then
requires $|\hat{B}_{E,p}|=\left(2\pi\right)^{-(d+1)/2}$. Let
us now use the WKB approximation. For normalizable solutions,
we have $C=0$, or $a=-ib$, so the normalization of the wavefunction
is fixed by
\begin{equation}
|b|=\nu^{\frac{1}{2}}z^{\frac{1}{2}+\nu}\left(2\pi\right)^{-\frac{d+1}{2}}\lim_{y\rightarrow0}y^{\nu}e^{S\left(y\right)}.
\end{equation}
The properly normalized WKB solution is then given by
\begin{equation}
\psi_{E,p}\left(\rho\right)=
\begin{cases}
\left(2\pi\right)^{-\frac{d+1}{2}}\nu^{\frac{1}{2}}z^{\frac{1}{2}+\nu}\left(U+\Delta U-E^{2}\right)^{-\frac{1}{4}}\lim_{y\rightarrow0}y^{\nu}e^{S\left(y\right)-S\left(\rho\right)}, &
\,\rho<\rho_{0};
\\
e^{i\frac{\pi}{4}}\left(2\pi\right)^{-\frac{d+1}{2}}\nu^{\frac{1}{2}}z^{\frac{1}{2}+\nu}\left(E^{2}-U-\Delta U\right)^{-\frac{1}{4}}\lim_{y\rightarrow0}y^{\nu}e^{S\left(y\right)}\left[e^{-i\Phi\left(\rho\right)}-ie^{i\Phi\left(\rho\right)}\right], &
\,\rho>\rho_{0},
\end{cases}
\end{equation}
 where $S\left(\rho\right)=\int_{\rho}^{\rho_{0}}d\rho^{\prime}\sqrt{U+\Delta U-E^{2}}$,
$\Phi\left(\rho\right)=\int_{\rho_{0}}^{\rho}d\rho^{\prime}\sqrt{E^{2}-U-\Delta U}$
and $\Delta U\equiv 1/\left(2\rho^{\prime}\right)^{2}$ (see appendix~\ref{sec:WKB}).

Using
this result, we can write our candidate smearing function as
\begin{equation}
K=e^{-\frac{d}{2}B}\int\frac{dE}{\left(2\pi\right)^{\frac{1}{2}}}\frac{d^{d}p}{\left(2\pi\right)^{\frac{d}{2}}}e^{i\left(E\left(t^{\prime}-t\right)-\vec{p}\cdot\left(\vec{x}^{\prime}-\vec{x}\right)\right)}\psi_{E,p}.\label{eq:K as FT}
\end{equation}
We recognize this integral as the inverse Fourier transform of $\psi_{E,p}$.
We will now show that this object does not exist%
\footnote{For a precise definition of what we mean by nonexistence, see
section~\ref{sec:modifyingbb}.}
because $\psi$ grows exponentially with momentum $p$.

First, let
E and $\rho$ be fixed. We then choose $p$ large enough so $\rho<\rho_{0}$,
i.e.\ so the $\rho$ we are considering is in the tunneling region. This choice is possible for any $\rho$. For concreteness, we can choose
\begin{equation}
p^{2}>E^{2}\rho^{k}.
\label{eq:pbound1}
\end{equation}
Then
\begin{equation}
\left|\lim_{y\rightarrow0}y^{\nu}e^{S\left(y\right)-S\left(\rho\right)}\right|=\lim_{y\rightarrow0}y^{\nu}\exp\left(\int_{y}^{\rho}d\rho^{\prime}\sqrt{\frac{\nu^{2}}{\left(\rho^{\prime}\right)^{2}}+\frac{p^{2}}{\left(\rho^{\prime}\right)^{k}}-E^{2}}\right),\label{eq:relevant integral}
\end{equation}
 and the integral is real-valued. Now let $0<\lambda<1$ such that
$y<\lambda\rho<\rho$ and split the integral accordingly:
\begin{equation}
\int_{y}^{\rho}=\int_{y}^{\lambda\rho}+\int_{\lambda\rho}^{\rho}\label{eq:split integral}.
\end{equation}
 Roughly speaking, the first integral provides the boundary data with
the correct asymptotic $y$-dependence, while the second integral
is responsible for the exponential behavior in $p$. In the first
integral, using \eqref{eq:pbound1}, we find
\begin{align}
\int_{y}^{\lambda\rho}d\rho^{\prime}\sqrt{\frac{\nu^{2}}{\left(\rho^{\prime}\right)^{2}}+\frac{p^{2}}{\left(\rho^{\prime}\right)^{k}}-E^{2}} & >\nu\log\left(\frac{\lambda\rho}{y}\right).\label{eq:bound1}
\end{align}
In the second integral, for $p$ large enough%
\footnote{For concreteness, choose e.g.\ $p^{2}>E^{2}\rho^{k}/(1-c^{2}).$}
we can find a constant $0<c<1$ such that
\begin{equation}
\int_{\lambda\rho}^{\rho}d\rho^{\prime}\sqrt{\frac{\nu^{2}}{\left(\rho^{\prime}\right)^{2}}+\frac{p^{2}}{\left(\rho^{\prime}\right)^{k}}-E^{2}}>\int_{\lambda\rho}^{\rho}d\rho^{\prime}\frac{cp}{\left(\rho^{\prime}\right)^{\frac{k}{2}}}=cz\rho^{\frac{1}{z}}\left(1-\lambda^{\frac{1}{z}}\right)p\label{eq:bound 2}.
\end{equation}
Putting everything together, we conclude that for $E$ and $\rho$
fixed, there exist $c,\lambda \in (0,1)$ and $p_{0}$ such that
\begin{equation}
\left|\lim_{y\rightarrow0}y^{\nu}e^{S\left(y\right)-S\left(\rho\right)}\right|>\left(\lambda\rho\right)^{\nu}\exp\left[cz\rho^{\frac{1}{z}}(1-\lambda^{\frac{1}{z}})p\right],\label{eq:lowerbound pure lifshitz}
\end{equation}
for all $p>p_{0}$. Hence the function $\psi_{E,p}$ grows exponentially
with $p$ and the smearing function defined in \eqref{eq:candidate K}
does not exist%
\footnote{This exponential behavior in $p$ is distinct from the behavior of $|B|/|b|$ in
$\alpha$ (see e.g.\ \eqref{eq:B/b large alpha}), since here we are interested in the
amplitude of the wavefunction at a fixed radial location $\rho$, and not its overall
normalization.}.

The inability to construct a smearing function is
due to the existence of trapped modes, which have to tunnel through
$V_{p}$ to reach the boundary. The boundary imprint of these modes
is suppressed by a factor of $e^{-cp}$, where $c$ is some positive
constant depending on the geometry. However, the normalization condition
\eqref{eq:boundary normalization} turns this suppression into an
exponential amplification: For any given mode the smearing function
takes the corresponding boundary data and amplifies it by an appropriate
factor to reconstruct bulk information. Consequently, trapped modes
receive a contribution $e^{+cp}$ in the smearing function integral.
As $p\rightarrow\infty$, the boundary imprint of trapped modes becomes
arbitrarily small, and as a result the smearing function integral diverges.

The splitting of the domain of integration into a near-boundary region
$[0,\lambda\rho]$ and a bulk region $[\lambda\rho,\rho]$ is crucial
for our proof: In the near-boundary region, we use the fact that no
matter how large $p$ is, we can make $\rho^{\prime}$ small enough such
that the cosmological- and mass-terms in the potential dominate over
$V_{p}$ and we can approximate $U\approx \nu^{2}/(\rho^{\prime})^{2}$.
Modes that tunnel through this part do not contribute an exponential
factor $\sim e^{p}$, but rather produce the correct boundary scaling
$y^{-\nu}$. This scaling is consequently stripped off by the $y^{\nu}$
factor in \eqref{eq:relevant integral}. In the bulk region near $\rho$,
however, there is a minimum value that $\rho^{\prime}$ can take,
so as we drive $p$ to infinity, eventually $U\approx p^{2}/(\rho^{\prime})^{k}$
becomes a very good approximation. This is what produces the exponential
factor in \eqref{eq:lowerbound pure lifshitz}.

We see that there are two qualitatively different limits of the potential:
$\rho\rightarrow0$ and $p\rightarrow\infty$. Both of them are important
for understanding the behavior of \eqref{eq:relevant integral}, which
is why we need to pick $0<\lambda<1$ to get a lower bound that reflects
this behavior. Simply setting $\lambda=0$ corresponds to approximating
$U\approx p^{2}/(\rho^{\prime})^{k}$ everywhere. However,
in doing so we would be neglecting the boundary scaling $y^{-\nu}$,
and consequently the lower bound \eqref{eq:lowerbound pure lifshitz}
would be zero. Similarly, $\lambda=1$ corresponds to approximating
$U\approx \nu^{2}/(\rho^{\prime})^{2}$ everywhere. While this
is certainly true for small $\rho^{\prime}$, we would be missing
the fact that the momentum part $V_{p}$ of the potential can still
dominate in any interval away from the boundary (i.e.\ close to $\rho$)
and lead to exponential growth. The bound \eqref{eq:lowerbound pure lifshitz}
would just be a constant independent of $p$ and we would not be able
to make the same conclusion about the smearing function.

\subsection{\label{sub:Phase-space-analysis}Momentum-space analysis}

\begin{figure}[t]
\centering
\includegraphics[height=6cm]{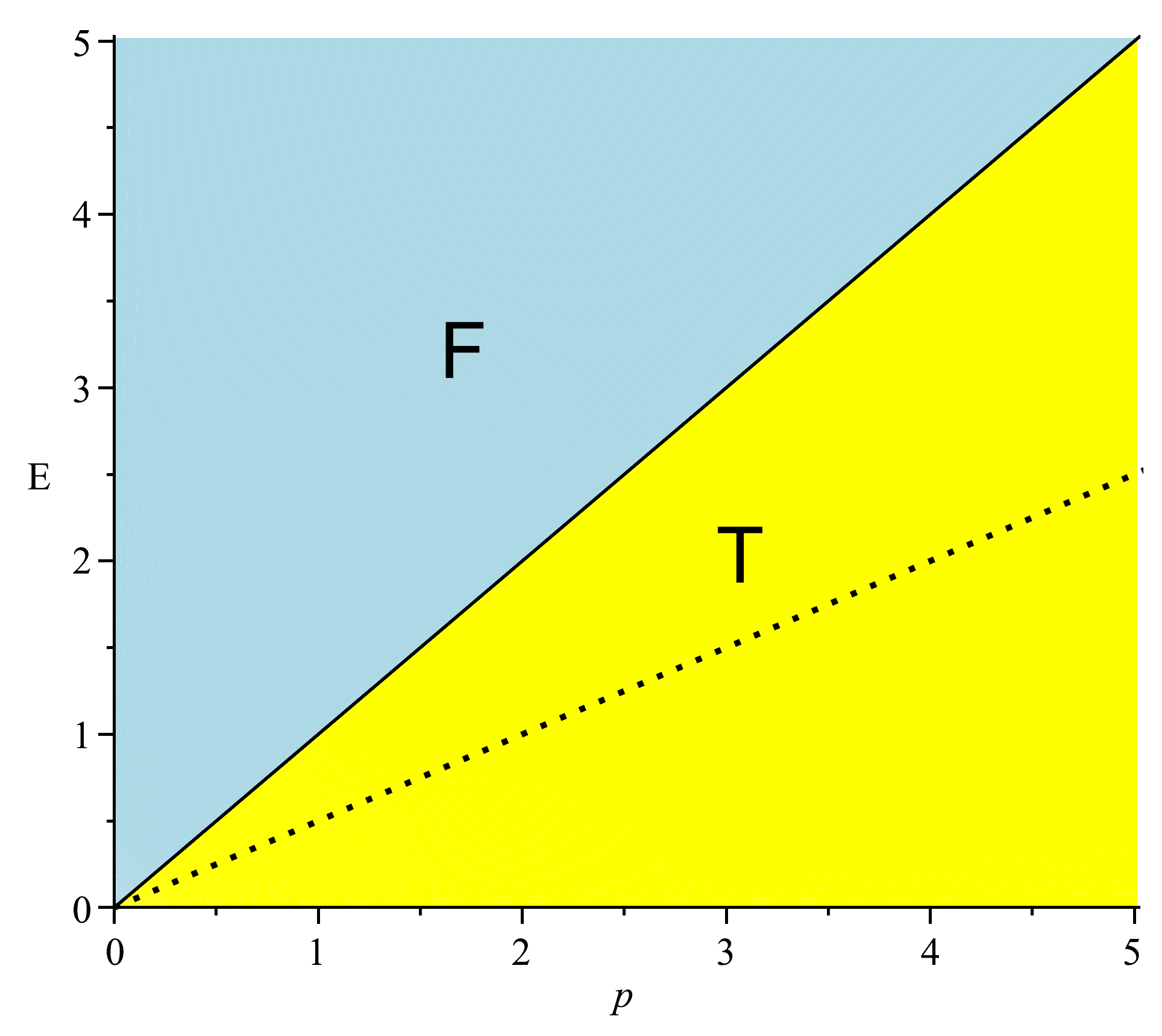}

\caption{\label{fig:E-p plane}Sketch of free (F) and trapped (T) modes for
general case. Deforming the geometry in the IR may introduce a cutoff
(dotted line), but this line will always remain below the solid line,
and some trapped modes survive.}
\end{figure}

It is instructive to analyze the behavior of the integral \eqref{eq:K as FT}
at large momenta in the (E,$|p|$)-plane. We already saw that for
fixed energy $E$, the smearing function diverges exponentially with
$|p|$, as the tunneling barrier becomes arbitrarily large at high momenta.
However, this is not necessarily the only direction along which the
integral diverges. Let us introduce polar coordinates
\begin{align}
|p| & =q\cos\theta\nonumber
\\
E & =q\sin\theta.
\label{eq:polar coords}
\end{align}
Figure~\ref{fig:E-p plane} shows a sketch of the spectrum in the
(E,$|p|$)-plane: The solid line divides trapped modes, which have
to tunnel through $V_{p}$ from `free' modes, which only tunnel through
$U\sim1/\rho^{2}$. If we imagine cutting off Lifshitz at
some small value $\lambda\rho$ with $\lambda<1$, all modes with
$E<\left(\lambda\rho\right)^{-\frac{1}{2}}|p|$ (yellow region) are
trapped modes%
\footnote{Notice that the choice of $\lambda$ is arbitrary. In particular,
along any line $E=\tan\theta|p|$, there is a choice of $\lambda$
such that all modes are below the momentum-barrier for large enough
$|p|$. Nevertheless, because of the subtleties discussed at the end
of the previous section, we should not simply take $\lambda\rightarrow0$
but instead work with a small but finite value.}.
Let us study the integral which defines the smearing direction. If we perform this integral along any direction $\theta$ over these
modes (i.e.\ $\tan\theta<\left(\lambda\rho\right)^{-\frac{1}{2}}$), the exponential term in the integrand behaves as
\begin{equation}
\mathrm{Re}\left(S\left(y\right)-S\left(\rho\right)\right)
=\int_{y}^{\rho}d\rho^{\prime}\sqrt{\frac{\nu_{z}^{2}}{(\rho^{\prime})^{2}}+\left(\frac{1}{(\rho^{\prime})^{k}}-\tan^{2}\theta\right)q^2\cos^{2}\theta }.
\end{equation}
For $q$ large enough, this term grows linearly and the smearing function
is exponentially divergent. We see that the variable that controls
the suppression (or amplification) due to tunneling is in fact $q=\sqrt{E^{2}+p^{2}}$,
as opposed to just $|p|$.
%

\subsection{\label{sub:adslifads}No smearing function $\Leftrightarrow$ singularities?}

The divergence of the smearing function is due to trapped modes, which
correspond to classical geodesics that cannot reach the boundary.
However, those are precisely the trajectories that start and end at
the tidal singularity at $\rho\rightarrow\infty$, so their fate is
not well-understood even on the classical level. Therefore, one might
wonder if the inability to construct smearing functions is simply
due to the presence of singularities. This question has been raised
before in the case of black hole solutions in AdS%
\footnote{However, we should point out that the two types of singularities encountered
here are qualitatively different. In the Lifshitz case, the singularity
is `mild', in the sense that all curvature invariants remain finite.
It is, however, felt by strings that fall towards the horizon \cite{Horowitz:2011gh}.%
} \cite{Bousso:2012mh,Leichenauer:2013kaa}. Fortunately, in our case
there are known ways to resolve the singularity, so we can directly
test the conjecture that non-existence of smearing functions is related
to singularities.

In the context of Einstein-Maxwell-dilaton systems \cite{Goldstein:2009cv},
the Lifshitz singularity can be resolved by including corrections
to the dilaton effective potential. For magnetically charged branes,
the dilaton runs towards strong coupling in the IR. Using a toy-model
of the quantum corrected action, the authors of \cite{Harrison:2012vy}
showed that the Lifshitz geometry can be resolved into an $\mathrm{AdS}_{2}\times\mathbb{R}^{2}$
region in the deep IR. For electrically charged solutions, the dilaton
runs towards weak coupling near the horizon, and higher derivative
corrections become important. In \cite{Knodel:2013fua}, two of the current authors showed
that by coupling the dilaton to higher curvature terms in an appropriate
way, the singularity can be resolved in a similar fashion. In particular,
numerical solutions were constructed that interpolate between $\mathrm{AdS}_{4}$
in the UV to Lifshitz in some intermediate regime, and finally to
$\mathrm{AdS}_{2}\times\mathbb{R}^{2}$ in the deep IR. We would like to use
these numerical flows to test whether resolving the singularity can
make the smearing function well-defined.

As a warm-up, consider the following analytical toy-model describing
such a flow:
\begin{align}
e^{2A} & =\frac{1}{\rho^{2}},\nonumber
\\
e^{2B} & =\begin{cases}
\frac{1}{\rho^{2}}, & 0<\rho< R_{1};
\\
\frac{1}{R_{1}^{k}\rho^{2-k}}, & R_{1}<\rho<R_{2};
\\
\frac{1}{R_{1}^{k}R_{2}^{2-k}}, & R_{2}<\rho,
\end{cases}\nonumber
\\
C & =A.\label{eq:adslifads toy model metric}
\end{align}
The last condition is a gauge choice, which fixes our radial coordinate
to be $\rho$, as defined in \eqref{eq:dxdr}. The potential is given
by
\begin{equation}
U\left(\rho\right)=\begin{cases}
\frac{\nu_{1}^{2}-\frac{1}{4}}{\rho^{2}}+p^{2}, & 0<\rho<R_{1};
\\
\frac{\nu_{z}^{2}-\frac{1}{4}}{\rho^{2}}+p^{2}\left(\frac{R_{1}}{\rho}\right)^{k}, & R_{1}<\rho<R_{2};
\\
\frac{\nu_{\infty}^{2}-\frac{1}{4}}{\rho^{2}}+p^{2}\left(\frac{R_{1}}{R_{2}}\right)^{k}\left(\frac{R_{2}}{\rho}\right)^{2}, & R_{2}<\rho,
\end{cases}\label{eq:adslifads toy model U}
\end{equation}
where $\nu_{z}$ was defined in \eqref{eq:nualphadef}, and $0<k<2$.
All modes with $p>E$, or equivalently $\tan\theta<1$ are trapped.
It is interesting to note that since the potential goes to zero as
$\rho\rightarrow\infty$, there are now modes that are below the barrier
in the $\mathrm{AdS}_{d+2}$ region. For pure AdS, this is not possible,
as the wavefunction cannot be below the barrier everywhere.

Let us see if a smearing function exists for any point $\rho$ in
the bulk. For $0<\rho<R_{1}$, we need to compute
\begin{equation}
\left|\lim_{y\rightarrow0}y^{\nu}e^{S\left(y\right)-S\left(\rho\right)}\right|
=
\lim_{y\rightarrow0}y^{\nu}\exp\left(\mathrm{Re}\int_{y}^{\rho}d\rho^{\prime}\sqrt{\frac{\nu_{1}^{2}}{\rho^{\prime2}}+\left(1-\tan^{2}\theta\right)q^2\cos^{2}\theta}\right).\label{eq:e^S in AdS}
\end{equation}
 Naively, one might expect that since we are integrating all the way
up to the boundary at $\rho=0$, the $1/\rho^{2}$-term will
eventually dominate and there is no $q$-divergence. However, we have
seen before that it is necessary to split the integral into a near-boundary
region and a bulk region, according to \eqref{eq:split integral}.
The near boundary integral will then produce the typical boundary
scaling $y^{-\nu}$, while the bulk integral will grow linearly for
trapped modes. In complete analogy with \eqref{eq:lowerbound pure lifshitz}
we find that there exist constants $q_{0},c>0$ and $\lambda\in(0,1)$
such that
\[
\left|\lim_{y\rightarrow0}y^{\nu_{1}}e^{S\left(y\right)-S\left(\rho\right)}\right|>\left(\lambda r\right)^{\nu_{1}}e^{cq},
\]
 for all $q>q_{0}$. Again, even though the $1/\rho^{2}$
part of the potential dominates near the boundary, there is still
an exponential divergence due to trapped modes, and the smearing function
does not exist in the AdS region.

For points within the Lifshitz region
($R_{1}<\rho<R_{2}$), the relevant integral contains an integral
over the $\mathrm{AdS}_{d+2}$ region, which is divergent by itself,
plus an additional term
\begin{equation}
\int_{R_{1}}^{\rho}d\rho^{\prime}\sqrt{\frac{\nu_{z}^{2}}{\rho^{\prime2}}+\left(\left(\frac{R_{1}}{\rho^{\prime}}\right)^{k}-\tan^{2}\theta\right) q^2 \cos^{2}\theta}.
\end{equation}
This integral gives a real contribution for $\tan\theta<\left(R_{1}/\rho\right)^{k/2}$,
which grows linearly with large $q$. Hence the smearing function
still grows like $e^{c^{\prime}q}$, but now $c^{\prime}>c$ and it
diverges even faster than in the $\mathrm{AdS}_{d+2}$ part.

The same
logic can be applied to a point within the $\mathrm{AdS}_{2}\times\mathbb{R}^{d}$
region in the IR ($\rho>R_{2}$). In this case there is a contribution
from both $\mathrm{AdS}_{d+2}$ and Lifshitz, plus a contribution
\begin{equation}
\int_{R_{2}}^{\rho}d\rho^{\prime}\sqrt{\frac{\nu_{\infty}^{2}}{\rho^{\prime2}}+\left(\left(\frac{R_{1}}{R_{2}}\right)^{k}\left(\frac{R_{2}}{\rho^{\prime}}\right)^{2}-\tan^{2}\theta\right)q^2 \cos^{2}\theta }.
\end{equation}
 Modes with $\tan\theta<\left(R_{1}/R_{2}\right)^{k/2}R_{2}/\rho$
begin to tunnel already in the $\mathrm{AdS}_{2}\times\mathbb{R}^{d}$
part of the potential, and so the smearing function will diverge even
faster at large $q$. The final result is that there is no smearing
function for any point $\rho$ in the bulk. The trapped modes lead
to an exponential divergence which becomes worse the deeper we try
to reach into the bulk.

\begin{figure}[t]
\centering
\includegraphics[height=6cm]{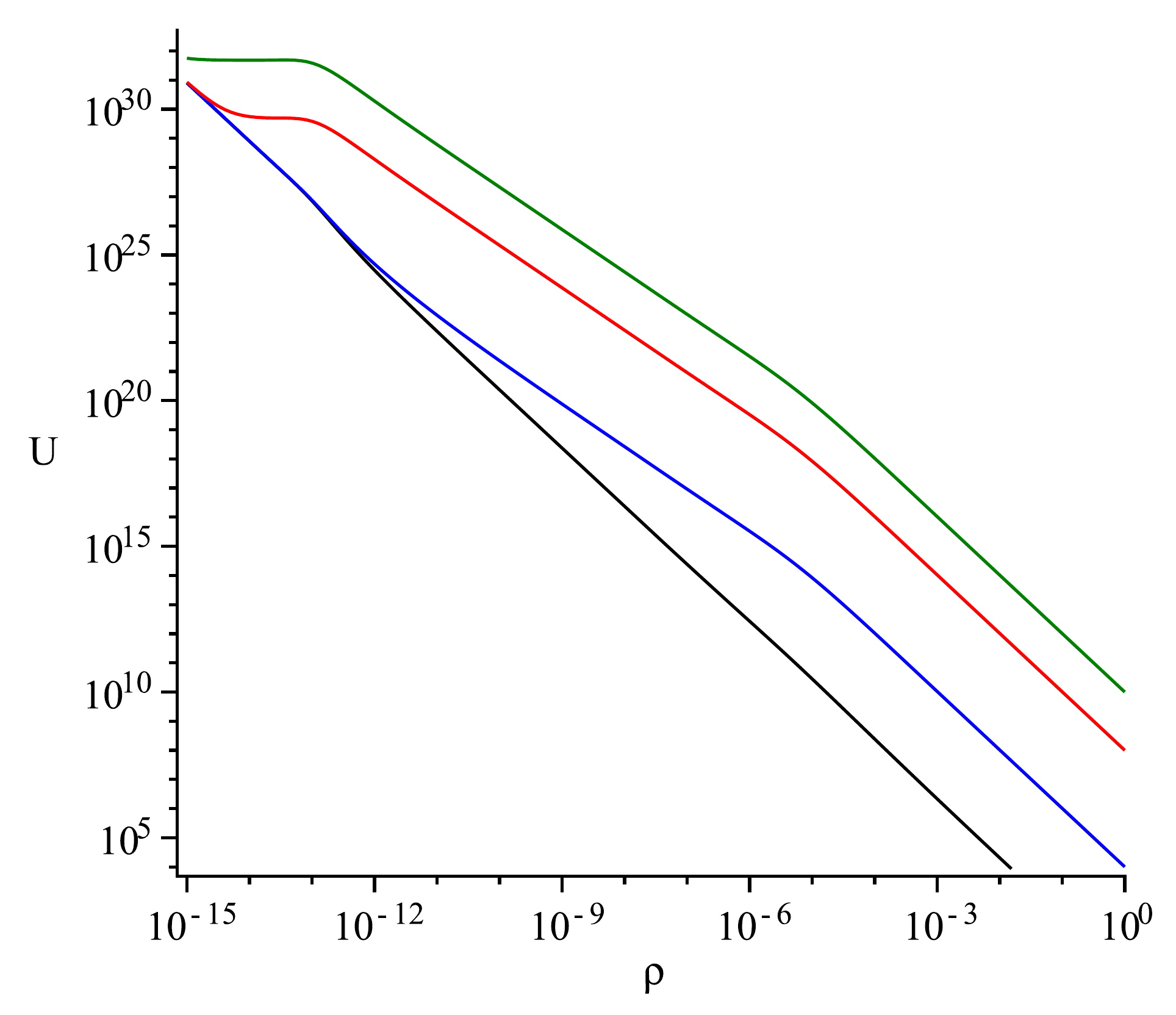}
\caption{\label{fig:U for adslifads}Effective potential $U$ for the numerical
flow found in \cite{Knodel:2013fua}, for $m=1$. The momentum increases from bottom to top,
with $p=0$ (black), $10^{2}$ (blue), $10^{4}$ (red), $10^{5}$ (green).
At large momenta, the potential is well approximated by $V_{p}=e^{2W}p^{2}$.}
\end{figure}

\begin{figure}[t]
\centering
\hspace{-0.5cm}\includegraphics[height=6cm]{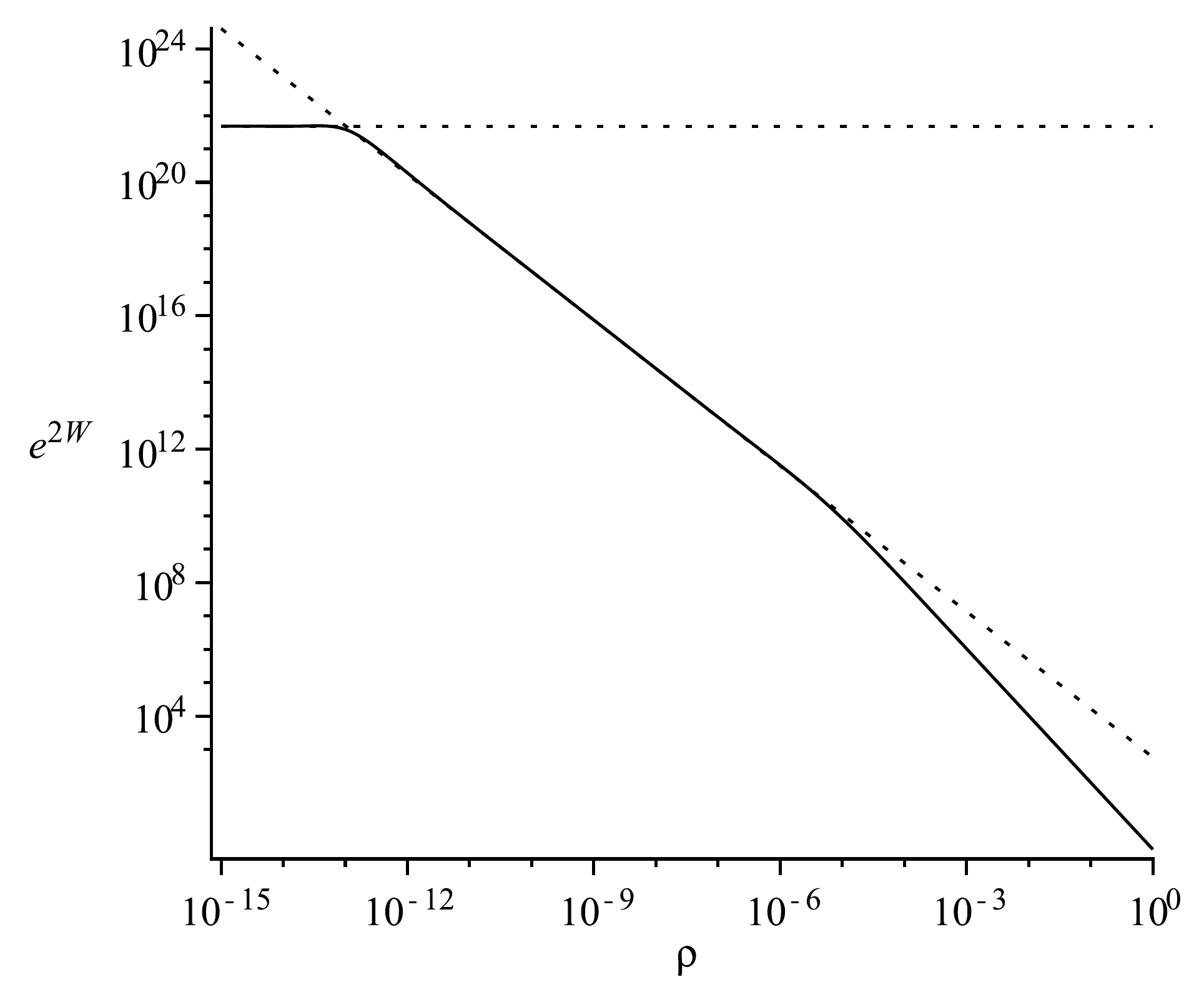}
\caption{\label{fig:adslifads Vp}The factor $e^{2W}$ for the same numerical solution. The
solution flows from $\mathrm{AdS}_{4}$ ($e^{2W}\approx \mathrm{const.}$),
to Lifshitz ($e^{2W}\sim\rho^{1.45}$, corresponding to $z\approx3.68$)
to $\mathrm{AdS}_{2}\times\mathbb{R}^{2}$ ($e^{2W}\sim\rho^{-2}$).}
\end{figure}

Let us now check that the result obtained for the toy-model \eqref{eq:adslifads toy model metric}
is indeed correct also for the exact numerical solution found in \cite{Knodel:2013fua}
(here $d=2$). The effective potential is plotted in Figure~\ref{fig:U for adslifads}.
As $p$ increases, the potential becomes better and better approximated
by $V_{p}$ (shown in Figure~\ref{fig:adslifads Vp}).
The metric coefficients and potential are of the form given in \eqref{eq:adslifads toy model metric}
and \eqref{eq:adslifads toy model U},
except that now there is a smooth transition between the three regions.

\begin{figure}[t]
\centering
\includegraphics[height=5.7cm]{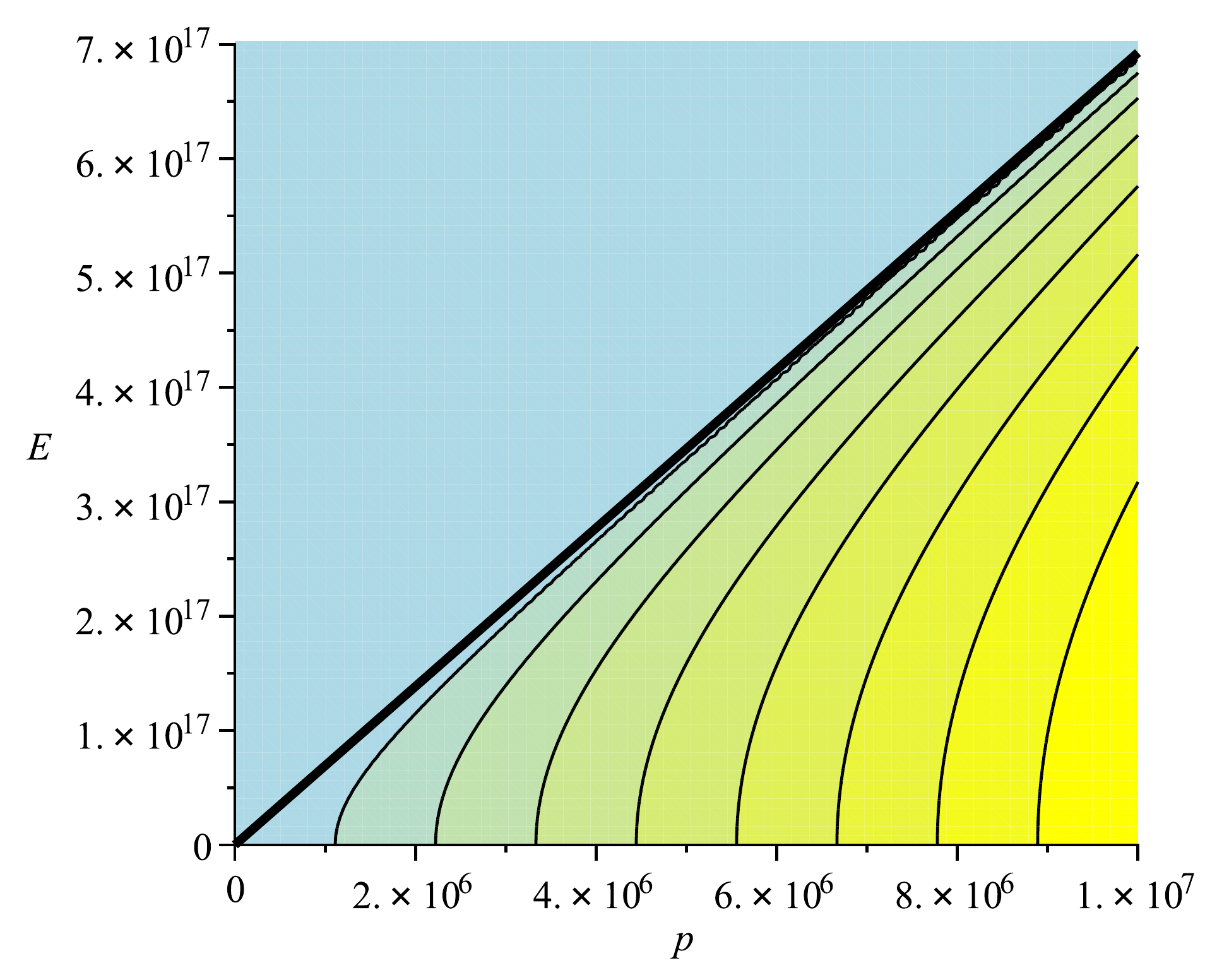}
\vspace{-.4cm}
\caption{\label{fig:ReSEpplaneAdS4}Plot of $\mathrm{Re\left(S(y)-S(\rho)\right)}$ for
a point within the $\mathrm{AdS}_{4}$ region ($\rho\approx1.3\cdot10^{-15}$).
The black solid line represents $V_{p}=E^{2}$ and divides free (blue)
from trapped modes (yellow). Contours indicate lines of constant
$\mathrm{Re\left(S(y)-S(\rho)\right)}$,
with a linear increase between different contours.}
\end{figure}

\begin{figure}[t]
\centering
\includegraphics[height=5.7cm]{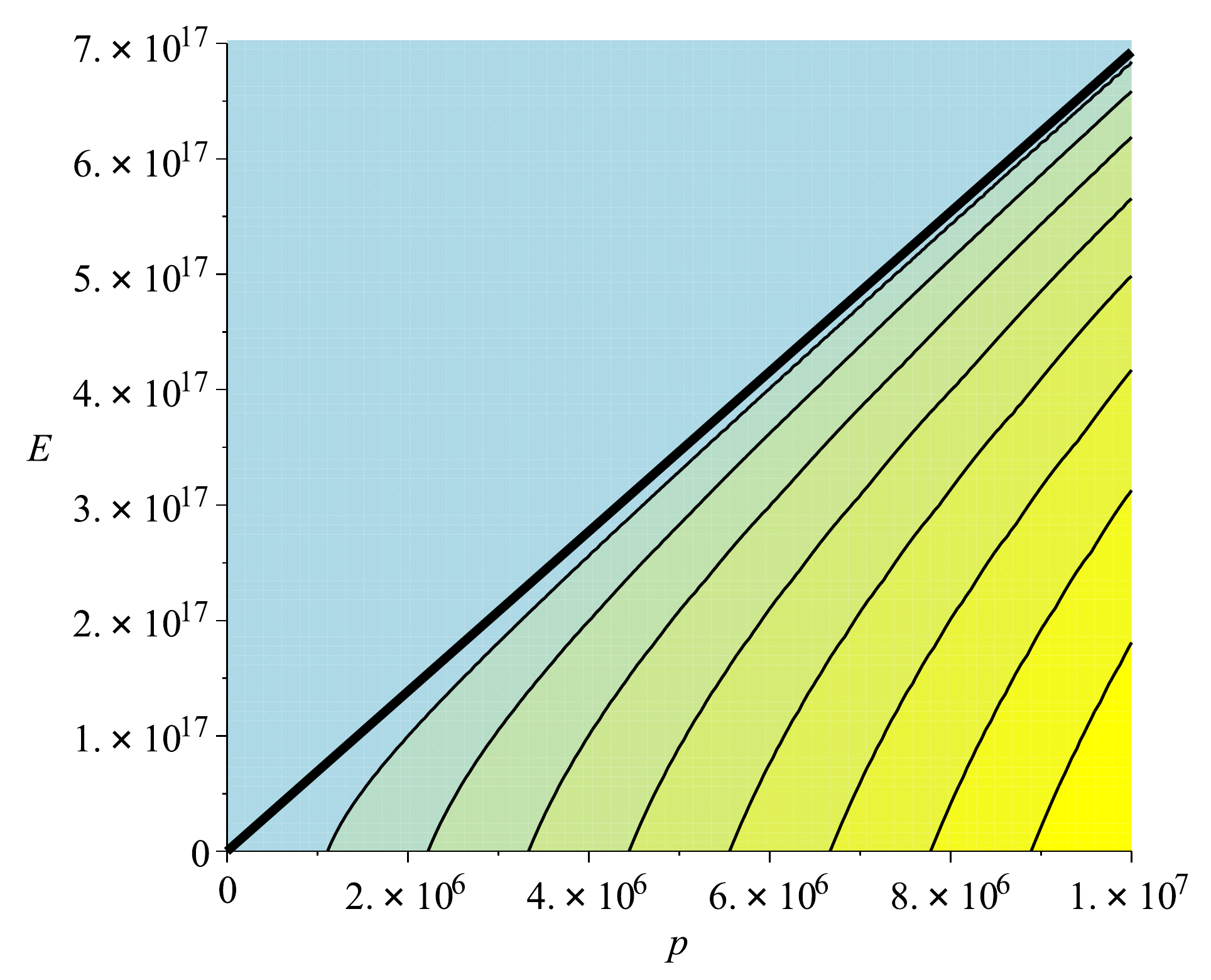}
\vspace{-.4cm}
\caption{\label{fig:ReSEpplaneLif}Plot of $\mathrm{Re\left(S(y)-S(\rho)\right)}$
for a point within the Lifshitz region ($\rho\approx9\cdot10^{-8}$).}
\end{figure}

\begin{figure}[t]
\centering
\includegraphics[height=5.7cm]{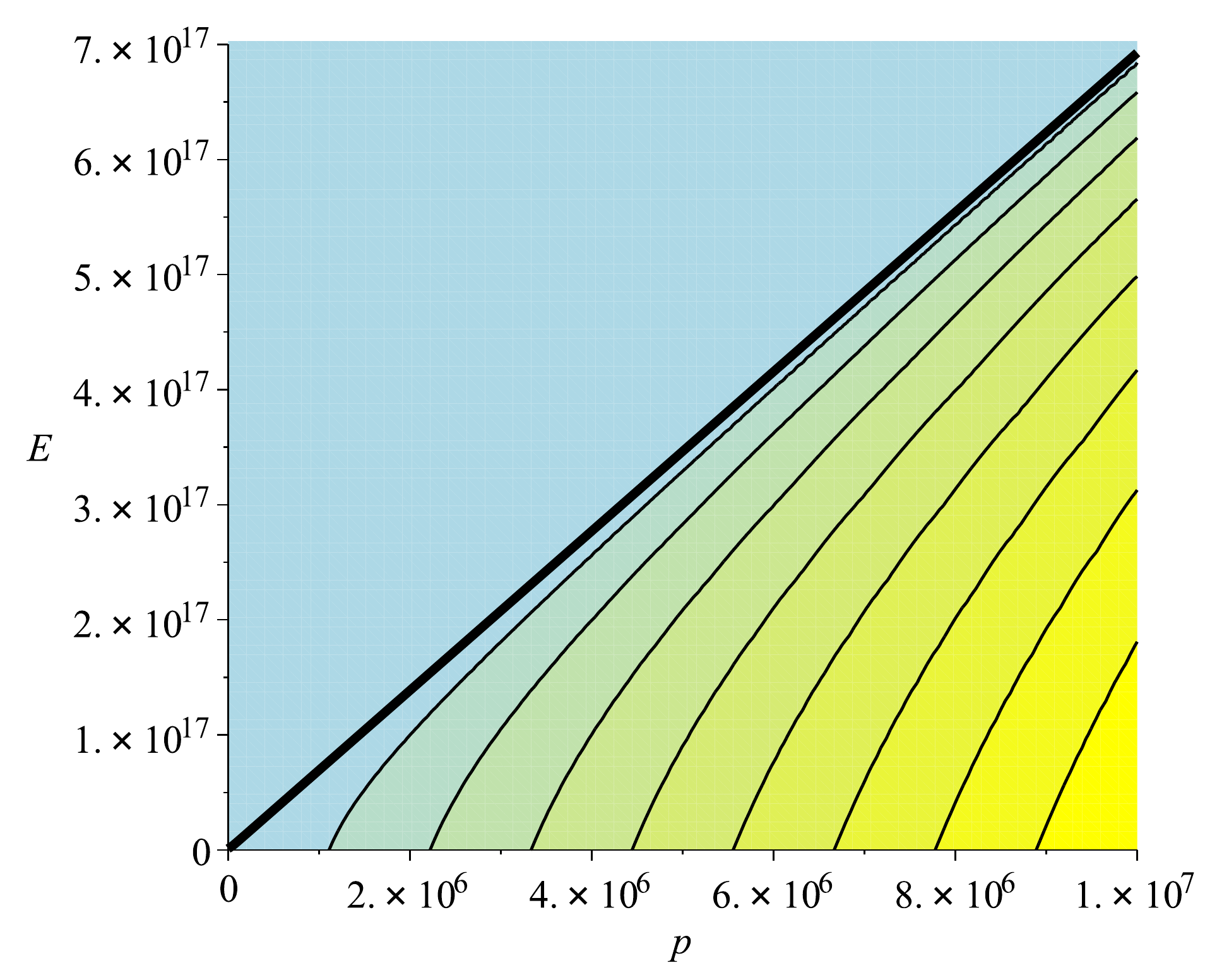}
\vspace{-.4cm}
\caption{\label{fig:ReSEpplaneAdS2}Plot of $\mathrm{Re\left(S(y)-S(\rho)\right)}$
for a point within the $\mathrm{AdS}_{2}\times\mathbb{R}^{2}$ region ($\rho\approx1$).}
\end{figure}

Figures~\ref{fig:ReSEpplaneAdS4}-\ref{fig:ReSEpplaneAdS2} show the
real part of $S\left(y\right)-S\left(\rho\right)$ in the (E,$|p|$)-plane.
Instead of taking $y$ to zero we choose $y\approx10^{-15}$, which
we may think of as disregarding the near-boundary region of the $\rho^{\prime}$-integral
and starting at $y=\lambda\rho$. The thick line divides free (blue)
modes from trapped (yellow) modes. The contours represent lines along
which $\mathrm{Re}\left(S(y)-S(\rho)\right)$ is constant. If we keep
$E$ fixed and increase $p$, we cross the contours at approximately
equal distances, so the integral grows linearly in $p$. This is not
only true for lines of constant $E$, but for any line within the
trapped region (i.e.\ any line that stays below the black solid line).
Hence the integral indeed diverges linearly with $q=\sqrt{E^{2}+p^{2}}$,
as was anticipated in section \ref{sub:Phase-space-analysis}.

\begin{figure}[t]
\centering
\hspace{-4cm}\includegraphics[height=7cm]{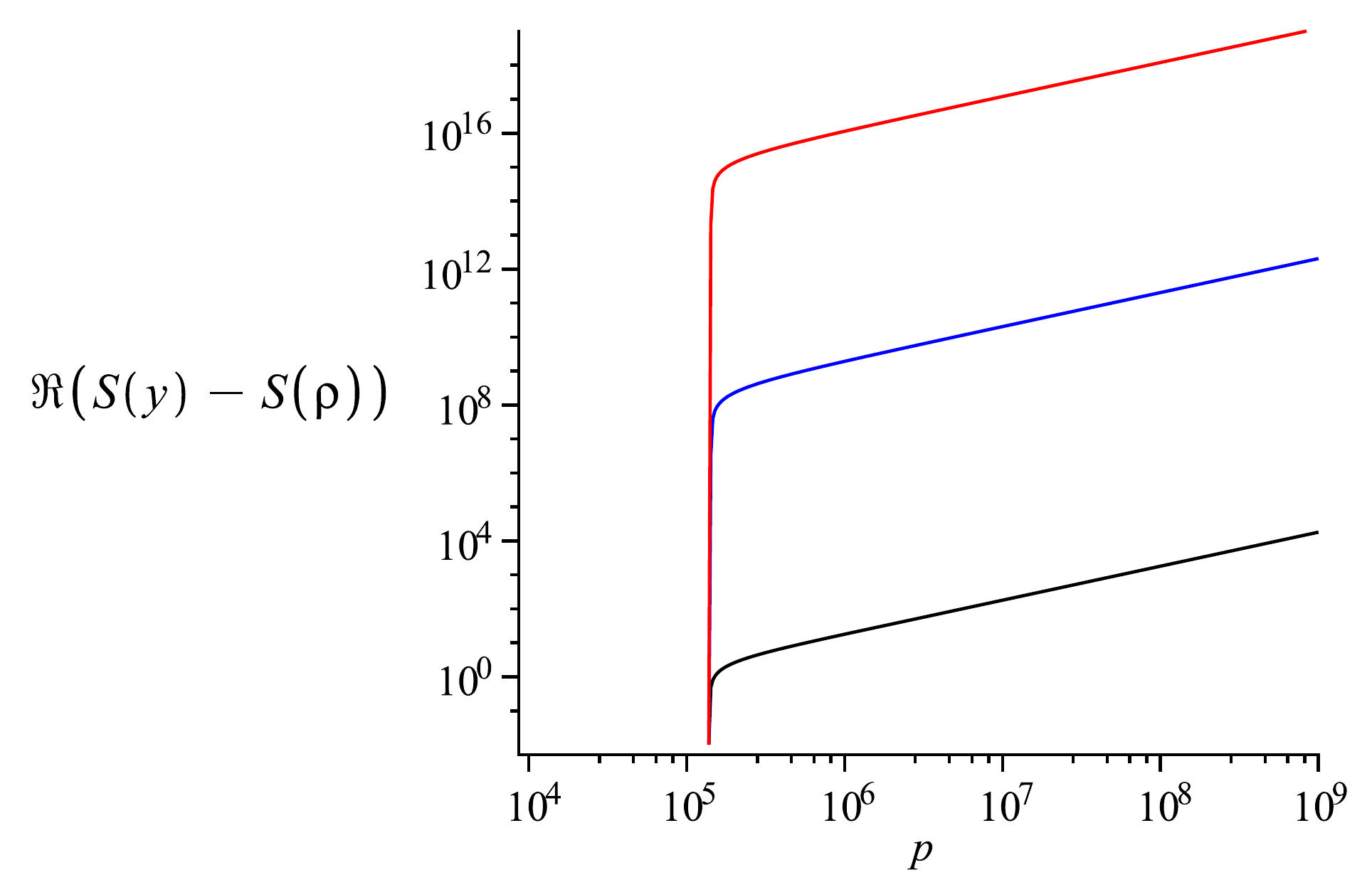}

\caption{\label{fig:Re(S) all points}Plot of the real part of $S(y)-S(\rho)$
vs. $p$ at three different positions within the $\mathrm{AdS}_{4}$ ($\rho\approx1.3\cdot10^{-15}$),
Lifshitz ($\rho\approx9\cdot10^{-8}$) and $\mathrm{AdS}_{2}\times\mathbb{R}^{2}$
($\rho\approx1$) regions (from bottom to top). The energy is fixed
at $E=10^{16}$ and we chose $m=1$. For large momenta, the solution
begins to tunnel and contributes an exponential factor in $K$.}
\end{figure}

Figure~\ref{fig:Re(S) all points} shows $\mathrm{Re}\left(S\left(y\right)-S\left(\rho\right)\right)$
for three points representing $\mathrm{AdS}_{4}$, Lifshitz and $\mathrm{AdS}_{2}\times\mathbb{R}^{2}$.
The energy is held fixed at $E=10^{16}$ , such that at small $p$,
the wavefunction is oscillating everywhere. As we increase $p$, the
mode eventually becomes trapped and the real part of the integral
grows linearly. Note that in the log-log plot used here, the three
curves lie nicely on top of each other. This fact confirms our prediction that
the smearing function diverges faster the deeper we try to reach into
the bulk.

We conclude that resolving the tidal singularity is not enough to
make the smearing function well defined. The $\mathrm{AdS}_{2}\times\mathbb{R}^{2}$
region in the IR can be thought of as the $z\rightarrow\infty$ limit
of Lifshitz spacetime. As a consequence, $V_{p}\sim\rho^{-2}$, and
there are still trapped modes with arbitrarily small boundary imprint.

It is also worth commenting on the addition of an AdS region in
the UV, as in \eqref{eq:adslifads toy model metric}, which may seem
desirable to make the holographic renormalization procedure better-defined.
We have seen explicitly that the integral over \eqref{eq:e^S in AdS}
is still divergent at large momenta and a smearing function does not
exist, even for points close to the boundary. This is the quantum
equivalent of the observation made at the end of section \ref{sub:lifgeos},
that null geodesics with large enough $p$ still see a `Lifshitz
barrier' and remain trapped inside the bulk, regardless of the near-boundary
geometry.

\subsection{\label{sub:lifads}Other flows involving Lifshitz}

The $\mathrm{AdS}_{2}\times\mathbb{R}^{d}$ geometry considered in the previous section is not the only possible IR endpoint of the RG-flow for Lifshitz solutions. Ref.~\cite{Braviner:2011kz,Singh:2010cj,Singh:2013iba} have considered flows from Lifshitz in the IR to an $\mathrm{AdS}_{d+2}$ fixed point in the UV. These flows are of particular interest to us, since $V_{p}$ does not go to zero as $\rho \rightarrow \infty$, but reaches a constant value corresponding to the AdS geometry at the horizon. Consequently, some of the problematic trapped modes never oscillate, and are thus removed from the spectrum. To see how this works, consider the following toy-model of such a Lifshitz to $\mathrm{AdS}_{d+2}$ flow:

\begin{align}
e^{2A} & =\frac{1}{\rho^{2}},\nonumber \\
e^{2B} & =\begin{cases}
\frac{1}{\rho^{2-k}}, & 0<\rho\leq R_{1};\\
\frac{R_{1}^{k}}{\rho^{2}}, & \rho>R_{1},
\end{cases}\nonumber \\
C & \equiv A.\label{eq:lifads4}
\end{align}
The potential is given by
\begin{equation}
U\left(\rho\right)=\begin{cases}
\frac{\nu_{z}^{2}-\frac{1}{4}}{\rho^{2}}+\frac{p^{2}}{\rho^{k}} ,& 0<\rho\leq R_{1};\\
\frac{\nu_{1}^{2}-\frac{1}{4}}{\rho^{2}}+\frac{p^{2}}{R_{1}^{k}} ,& \rho>R_{1}.
\end{cases}
\end{equation}
To compute the smearing function at some fixed $\rho\leq R_{1}$
we again split the interval {[}0,$\rho${]} into a near-boundary region
$[0,\lambda\rho]$ and a bulk region {[}$\lambda\rho$, $\rho${]},
where $\lambda<1$. In the bulk region, the potential can be approximated
by $V_{p}=p^{2}/\rho^{k}$ for $p$ large enough. Then, modes
with $p>\left(\lambda\rho\right)^{k/2}E$ are trapped by $V_{p}$.
For $\rho>R_{1}$, the potential takes a constant value. In pure
Lifshitz, modes with $p<R_{1}^{k/2}E$ would have been
oscillating in this region. However, these modes are now completely
under the barrier and therefore have to be excluded from the spectrum.
The $\mathrm{AdS}_{d+2}$ region in the IR thus introduces a natural (energy-dependent)
momentum cutoff.

Nevertheless, there is still a finite wedge of trapped modes with
$R_{1}^{-k/2}<\tan\theta<\left(\lambda\rho\right)^{-k/2}$
(cf. Figure~\ref{fig:E-p plane}) and integrating up to $q=\infty$
will produce the same divergent behavior as before. In section \ref{sub:removing trapped modes},
we will give a general argument as to why this has to be the case, and
show that no smooth IR-deformation can remove all trapped modes from
the spectrum.

\section{\label{sec:sfgeneral}Generalization}

We have seen that the construction of smearing functions can fail
if there are modes that have to tunnel through a momentum barrier
in the potential. The integral \eqref{eq:candidate K} diverges if
such modes exist at arbitrarily large $q=\sqrt{E^{2}+p^{2}}$. In
this section, we will generalize our previous findings to prove that
smearing functions do not exist for any geometries that allow trapped
modes.

Consider a background that satisfies
\begin{equation}
\partial_{\rho}e^{W}<0\;\mathrm{for}\;\rho\in[\rho_{1},\rho_{2}].\label{eq:preliminary crit}
\end{equation}
We would like to compute the smearing function at a bulk point $\rho>\rho_{1}$.
All modes with $V_{p}\left(\rho_{1}\right)>E^{2}$ have to tunnel
through some part of $V_{p}$ and are therefore trapped modes. Let
us write the integral defining the smearing function in \eqref{eq:candidate K} as $\int dEd|p|\int d\Omega_{d-1}$
and focus on the integral in the ($E$,$|p|$)-plane. The domain of
integration is shown in Figure~\ref{fig:E-p plane}, where free and
trapped modes are separated by the solid line $E^{2}=V_{p}\left(\rho_{1}\right)$.
Choosing polar coordinates \eqref{eq:polar coords}, we find that the exponential part of the integrand satisfies
\[
\mathrm{Re}\left(S\left(y\right)-S\left(\rho\right)\right)>\mathrm{Re}\int_{\rho_{1}}^{\rho_{2}}d\rho^{\prime}\sqrt{V_{\mathrm{m}}(\rho^{\prime})+V_{\mathrm{cos}}(\rho^{\prime})+\left(e^{2W(\rho^{\prime})}-\tan^{2}\theta\right)\cos^{2}\theta q^{2}}
\]
Since the integration domain does not include the boundary, the first
two terms under the square root are bounded. Thus, for $\tan\theta<e^{W(\rho_{1})}$, the
integral grows linearly with large $q$ and the smearing function
diverges exponentially. The divergence appears not only at fixed $E$,
but under any angle in the yellow region of Figure~\ref{fig:E-p plane}.

Consequently, if a geometry has trapped modes that are below the barrier
at some $\rho_{1}$, a smearing function does not exist for any $\rho>\rho_{1}$.
From the null energy condition \eqref{eq:nullxxsimple} and the discussion
thereafter, we know that once $\partial_\rho e^{W}$ is negative for some $\rho_1$,
it cannot be positive for
any $\rho<\rho_{1}$.  Thus, once the wavefunction is below the $V_{p}$
barrier, it will stay below it as we go towards the boundary. Using
the terminology introduced in section \ref{sec:classical}, trapped
modes cannot become free near the boundary. Therefore, when computing
the smearing function $K\left(t,x,\rho|t^{\prime},x^{\prime}\right)$,
there is an exponential contribution from trapped modes regardless
of which bulk point $\rho$ we consider.

The condition \eqref{eq:preliminary crit} makes it is easy to identify
geometries without smearing functions. Clearly, Lifshitz has $\partial_\rho e^{W}<0$
everywhere, and as we saw earlier, $K$ does not exist. If we instead consider
flows that involve only a finite region with broken Lorentz invariance,
such that \eqref{eq:preliminary crit} is satisfied in some region, we still have trapped modes, and the smearing function will not exist.
This analysis includes flows involving a Lifshitz region, as well as hyperscaling geometries with Lifshitz scaling.
Our analysis above shows that none of these geometries admit smearing
functions, provided the spacetime satisfies the NEC.

\subsection{\label{sub:removing trapped modes}Removing trapped modes via deformations}

In our discussion above, we always assumed that the momentum-space integral
\eqref{eq:candidate K} does in fact include trapped modes with arbitrarily
large $q$ on some set of nonzero measure. This is clearly the case
in the examples mentioned above. On the other hand, the smearing function
for $\mathrm{AdS}$ converges because modes with $p^{2}>E^{2}$ are
simply not part of the spectrum, as the corresponding wavefunction
would have to be below the potential globally.

One might wonder if it is possible to `fix' a geometry which a
priori does not admit a smearing function, by removing all trapped
modes from the spectrum in a physical way. The AdS example gives us
a hint on how one might accomplish this task: If the geometry is deformed
in the deep IR such that would-be trapped modes never actually oscillate,
they would simply not be allowed.

\begin{figure}[th]
\noindent \begin{centering}
\includegraphics[height=6cm]{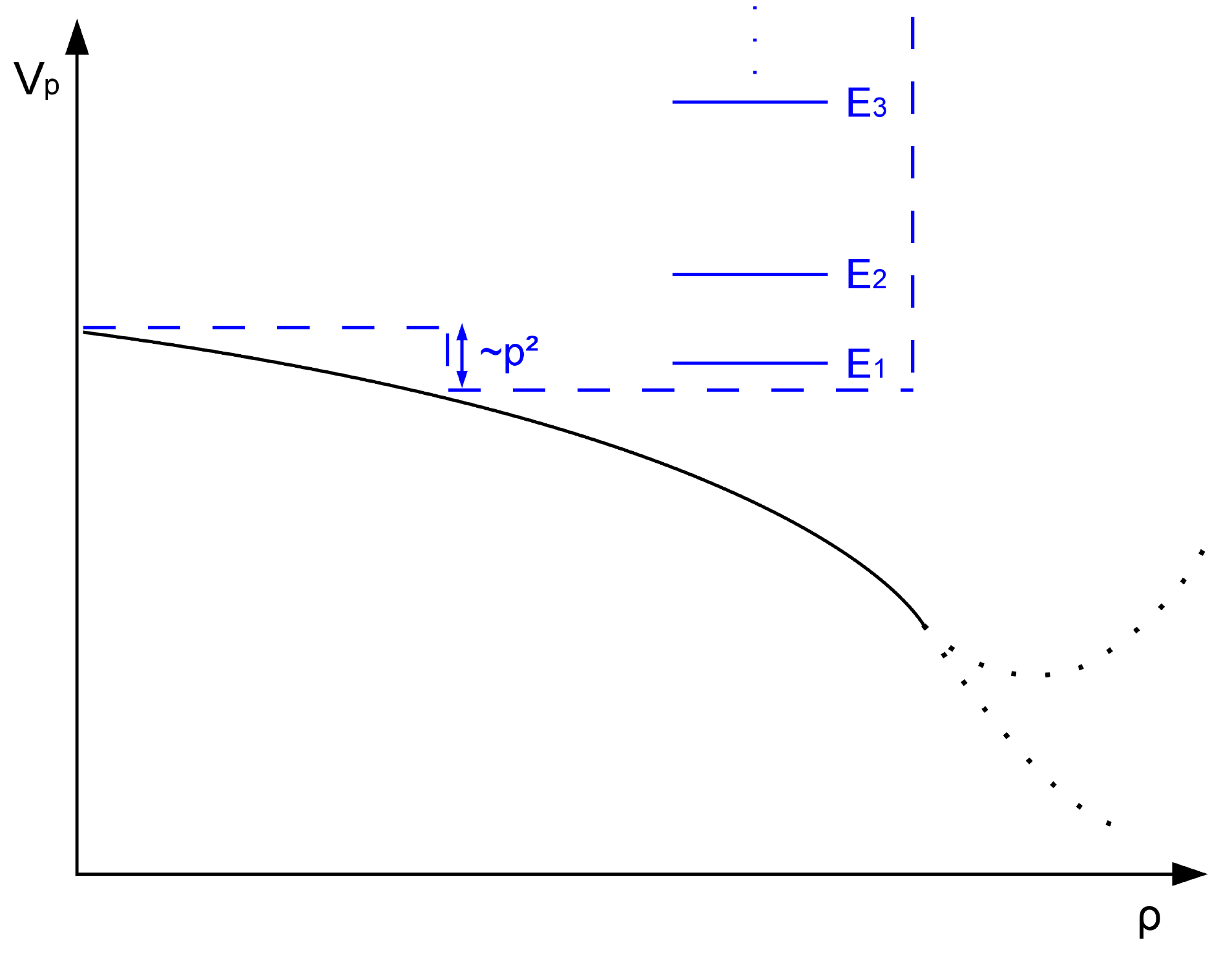}
\par\end{centering}

\caption{\label{fig:minmax}Sketch of $V_{p}$ for a potential satisfying \eqref{eq:preliminary crit}.
This includes deformations of AdS and flows involving Lifshitz. Using
the min-max principle, the energy levels are bounded from above by
those of a square-well potential. In the large $p$ limit, there are
always trapped modes. The near-horizon behavior of the potential is
irrelevant for our discussion. }
\end{figure}

Following our discussion of the
null energy condition in section \ref{sec:classical}, it follows
that there are only three relevant IR asymptotics that we need to
consider:
\begin{enumerate}
\item $e^{W}$ decreases monotonically to a constant value $\m>0$.
\item $e^{W}$ attains a minimum value $\m>0$, but then goes to constant
$M>\m$.
\item $e^{W}$ attains a minimum value $\m>0$, but then goes to infinity.
\end{enumerate}
Trapped states are equivalent to tunneling states in the potential
$V_{p}=p^{2}e^{2W}.$ For $p$ large enough, these states always exist
\cite{Brau:2003}. This can be seen heuristically by bounding the
potential from above with an appropriate square-well potential $\widetilde{U}\left(\rho\right)$
(see Figure~\ref{fig:minmax}). Therefore, no smooth deformation can
ever remove all trapped modes from the spectrum.

As an example, consider case 1, which captures the case of the Lifshitz
to $\mathrm{AdS}_{d+2}$ flow discussed in section \ref{sub:lifads}.
The AdS region introduces an energy-dependent momentum cutoff $p<E/\m$.
However, since $\m$ is by definition a global minimum and \eqref{eq:preliminary crit}
holds, we clearly have $\m<e^{W(\rho_{1})}$. Although the cutoff may
remove some trapped modes from the spectrum, there will always remain
a wedge of trapped modes that gives a divergent contribution when
integrated over (see Figure~\ref{fig:E-p plane}). We conclude that
spaces without a smearing function cannot be deformed smoothly to make
the smearing function well-defined.

\subsection{Adding trapped modes via deformations}

Another interesting question is what happens if we take a geometry
with a smearing function, such as AdS \cite{Bena:1999jv,Hamilton:2005ju,Hamilton:2006az},
and add a small (planar) perturbation in the IR.
It can be seen from \eqref{eq:nullxxsimple} that
$e^W$ must start with non-positive slope at the boundary for any background that
is asymptotically AdS%
\footnote{
If we do not insist on AdS asymptotics, then
we could choose $e^W$ to immediately have a positive slope. If $e^{W}$ has positive slope at some
$\rho_{+}$, the NEC dictates that $e^W$ cannot begin to decrease at
some larger $\rho$. Thus, in this scenario no trapped modes are
introduced, and the smearing function will continue to exist everywhere.  In particular, we cannot have a situation akin to Figure 5 in \cite{Leichenauer:2013kaa}, where the potential has a dip allowing trapped modes to become oscillating again close to the boundary.
}.
Since the potential scales with $p$, such a perturbation will always introduce new trapped
modes.  In particular, the momentum-potential $V_{p}=p^{2}e^{2W}$
can always be bounded from above by a semi-infinite square-well potential
of width $l$ and height $h=p^{2}h_{0}$, where $h_{0}$ is some constant
(see Figure~\ref{fig:minmax}). For large enough $p$, the square-well
always admits bound states with $p^{2}\left(1-h_{0}\right)<E^{2}<p^{2}$
and, via the min-max principle, so will $V_{p}$.  As a result, the smearing function would
be destroyed anytime the metric is deformed by such a perturbation.

This result is interesting, as it opens up the possibility that `small' perturbations of AdS can make the smearing function ill-defined by introducing new trapped states. However, we should keep in mind that our ansatz only allows for planar perturbations; we cannot consider localized disturbances. It would be interesting to study the effect of such perturbations in a more general setup. Again, notice that the ultimate IR fate of the geometry with AdS behavior in the UV is not important for this discussion.
In particular, whether or not there is a singularity at $r\rightarrow\infty$
does not change the qualitative result.

\subsection{Relativistic domain wall flows}

Given the above considerations, one may get the impression that the smearing function no
longer exists for any geometry other than pure AdS.  However, it is important to realize that
such a conclusion is in fact unwarranted.  What we have seen is that the non-existence of
the smearing function is intimately tied to the presence of trapped modes with exponentially
small imprint on the boundary.  Since such modes arise from the large $p$ limit of
$V_p=e^{2W}\vec p\,^2$, they are naturally absent when $W=0$, corresponding to flows
preserving $(d+1)$-dimensional Lorentz symmetry
\begin{equation}
ds_{d+2}^2=e^{2B(r)}[-dt^2+d\vec x_d^2]+e^{2C(r)}dr^2.
\end{equation}
In this case, the Schr\"odinger equation \eqref{eq:schr} is more naturally written as
\begin{equation}
-\psi''+(V_m+V_{\rm cos})\psi=(E^2-\vec p\,^2)\psi.
\label{eq:relsch}
\end{equation}
In particular, the effective potential $\hat U=V_m+V_{\rm cos}$ no longer scales with $p$.

In general, $\hat U$ may admit bound states and/or modes trapped at the horizon.  Although
bound states fall off exponentially outside the classically allowed region, since such states
occur only at fixed values of $Q^2\equiv E^2-\vec p\,^2$, they will always have a non-vanishing
(although small) amplitude at the boundary.  Hence the presence of such states do not
present an obstruction to the existence of a smearing function.  Trapped modes at the horizon,
on the other hand, are potentially more troubling, as they may form a continuum spectrum
with a limit of vanishing amplitude on the boundary.  However, it turns out that this
possibility does not prevent the construction of a well-defined smearing function
$K(t,x,r|t',x)$ for any fixed value of $r$.  The point here is that since $\hat U$ is independent of
$Q$, the maximum suppression factor to tunnel from the boundary to $r$ is bounded by setting
$Q=0$ in \eqref{eq:relsch}.  As a result, it is impossible to make the suppression arbitrarily
small.  Hence we conclude that the smearing function exists for finite $r$ in the case of
relativistic domain wall flows, although the $r\to\infty$ limit of $K$ may not exist if there are
trapped modes that live arbitrarily far from the boundary.

We see that it is generally possible to define a smearing function only for
relativistic flows, where $W=0$ along the entire flow.  Furthermore, for the case of
$\mathrm{AdS}_{d+2}\to \mathrm{AdS}_{d+2}$ flows, the effective potential $\hat U$
falls off as $1/\rho^2$ both in the UV and the IR.  Since this potential is too steep to admit
trapped modes in the deep IR, there are no modes completely removed from the boundary,
and hence the $r\to\infty$ limit of the smearing function is well-defined.  Thus in this case
the entire bulk may be reconstructed.

\section{\label{sec:modifyingbb}Modifying the bulk-boundary dictionary }

We have seen that for transverse Lorentz-breaking spacetimes with locally decreasing
transverse speed of light, the smearing function is not well defined,
even after resolving potential singularities. Thus, we are left with
the option of loosening some of our initial assumptions about this
function and its corresponding entry in the bulk-boundary dictionary.
In particular, we need to reexamine our implicit assumption that $K$
can reconstruct the bulk up to arbitrarily small transverse length
scales.

Let us be a bit more precise about what kind of mathematical object
the smearing function really is, and what we mean by saying that $K$
does or does not exist. The most general possible definition is to
let the smearing function be any map from boundary operators to bulk
fields. However, a reasonable condition is that $K$ defines a continuous,
linear functional on the space of boundary operators. Continuity means
that for any convergent sequence of boundary operators $O_{n}$ we
have
\begin{equation}
\lim_{n\rightarrow\infty}K\left[O_{n}\right]=K\left[\lim_{n\rightarrow\infty}O_{n}\right].\label{eq:continuity}
\end{equation}
 The difficulty in constructing such a $K$ is due to the fact that
the two limits are defined with respect to very different norms. The
bulk norm relevant for the left hand side is the Klein-Gordon norm \eqref{eq:KGNorm},
while the boundary norm for $O$ is given by \eqref{eq:boundary normalization}.
We have seen that in spacetimes with $\partial_\rho e^{W}<0$
locally, there exist nonzero bulk solutions that have exponentially
small boundary imprint, which provide an obstruction for constructing
continuous smearing functions.

Our strategy in this paper was to calculate a candidate smearing function
$\widehat{K}$ in momentum space, and ask whether it defines a well-behaved
object in position space. The problematic case is when the function
defined in this way grows exponentially, i.e.\ $\widehat{K}\approx e^{cp}$.
Its action on a boundary field can be written in momentum space as
\begin{equation}
K\left[O\right]\sim\int dp\,\widehat{K}\left(p\right)\widehat{O}\left(p\right)\label{eq:K(O)pspace}.
\end{equation}
 Whether or not this integral is well-defined clearly depends on what
we allow $\widehat{O}$ to be: If $\widehat{O}$ is a square-integrable
function, the smearing function has to be square-integrable as well,
which is clearly not the case here.

What if we impose a stricter fall-off
condition at $p\rightarrow\infty$? One rather strict condition would
be that $\widehat{O}$ falls off faster than any inverse power of
$p$ at infinity%
\footnote{In other words: $O$ is a Schwartz-function and $K$ is a tempered
distribution. %
}. A classic example of such a function is a Gaussian $\sim e^{-p^{2}}$.
However, $e^{cp}$ is not a well-defined functional on this space
either. This can be seen by explicitly constructing a sequence of
functions with `arbitrarily small' boundary imprint, i.e.\ a sequence
that goes to zero in the boundary norm. For example, consider
\begin{equation}
\widehat{O}_{n}\left(p\right)\equiv e^{-cn}\Psi\left(p-n\right),
\end{equation}
where $\Psi$ is some bump-function. Attempting to reconstruct the
corresponding bulk solution yields $K[O_{n}]\sim\int dp\Psi\left(p\right)$,
which is independent of $n$, and in particular never equal to zero.
Using \eqref{eq:continuity}, this means that the smearing function
is not continuous.

The only way to make sense of the smearing function
is to completely avoid configurations with arbitrarily small
boundary imprint. This can only be achieved by introducing a hard momentum
cutoff $\Lambda$. In other words, we attempt to invert the bulk-boundary
map $\phi\mapsto O$ only for configurations with $\widehat{O}(p>\Lambda)=0$.
Acting on these functions, the exponential $e^{cp}$ is indeed a well-defined
continuous functional, and the integral \eqref{eq:K(O)pspace} converges.
There is, however, a price to pay: as is well-known, the Fourier transform
of such compactly supported functions does not have compact support.
The position space wavefunction necessarily has to `leak out' to infinity,
and thus full localization in the transverse direction can never be
achieved%
\footnote{Here we have taken the necessity of smearing $\phi$ in position
space as an indication of nonlocality.  However, from a quantum point of view,
a more proper indication of nonlocality would be the nonvanishing of the commutator
outside of the lightcone.}.
%

\section{\label{sec:conclusion}Conclusion}

Motivated by some of the difficulties that have been observed in trying to understand the
global structure of Lifshitz spacetimes, we have studied the possibility of bulk reconstruction
from the boundary information.  At the classical level, the presence of non-radial
null geodesics that do not reach the Lifshitz boundary suggests that much of the bulk
data is inaccessible from the boundary.  We have confirmed this heuristic picture by
studying smearing functions for a bulk scalar field and demonstrating that they do not
exist for Lifshitz spacetimes with $z>1$.  The reason for this is that there will always
be trapped modes in the bulk that have exponentially vanishing imprint on the boundary.
It is these modes and the information that they contain that cannot be reconstructed
from any local boundary data.

Of course, it is well known that a pure Lifshitz background has a tidal singularity at the
horizon.  Since the trapped modes begin and end in the tidal singularity, we had initially
conjectured that resolving the Lifshitz singularity would remove such modes and lead to
a well defined smearing function.  However, this is not the case, as we have seen; even
with a regular horizon such as $\mathrm{AdS}_2\times \mathbb R^d$ or
$\mathrm{AdS}_{d+2}$, there will be trapped modes with vanishing imprint on the boundary
as the transverse momentum is taken to infinity.  Thus the existence or non-existence of a
smearing function is independent of the nature of the horizon, and in particular whether
it is singular or not.

More generally, we have seen that the constructibility of the smearing function depends
crucially on whether there exists a family of trapped modes with arbitrarily small
suppression on the boundary.  The only way this can arise is if the momentum
dependent part of the effective Schr\"odinger potential $V_p=e^{2W}\vec p\,^2$
has a local minimum or a barrier that grows as $p\to\infty$.  Thus the question of whether
the smearing function exists is closely related to the behavior of the gravitational redshift
factor $e^{-W}$.  In general, all non-relativistic backgrounds such as Lifshitz and ones
with hyperscaling violation (including flows with such regions) do not admit smearing
functions.  The same is true for geometries such as Schwarzschild-AdS, where $e^{2W}$
starts out as unity on the boundary, but vanishes at the horizon \cite{Leichenauer:2013kaa}.  On the other hand, smearing
functions are expected to exist for backgrounds with $W=0$, i.e.\ ones preserving $(d+1)$-dimensional
Lorentz invariance along the entire flow.

The scaling of $V_p$ with $\vec p\,^2$ has the important consequence that any trapped mode
will always be completely suppressed on the boundary with a factor $\sim e^{-cq}$ as
$q\to\infty$, where $q^2=E^2+\vec p\,^2$ and $c$ is a geometry and radial location dependent
positive constant.  This gives rise to the perhaps somewhat unexpected feature that, with
the existence of trapped modes, the smearing function $K(t,x,r|t',x)$ cannot exist even in an
asymptotic AdS$_{d+2}$ region near the boundary, so long as $r$ is at a fixed location.
One may wonder why the presence of trapped modes living in the IR would destroy the
possibility of reconstruction of the UV region near the boundary.  The reason for this is
that, while a trapped mode in the IR indeed has to tunnel to reach the boundary, its amplitude
does not immediately vanish in the interior of the bulk geometry.  Moreover, these modes
can live at a finite distance from the boundary.  Hence they can have an imprint at any
fixed $r$ in the bulk, and yet vanish on the boundary.  It thus follows that the bulk information
corresponding to such modes cannot be obtained from the boundary, and thus the smearing
function would not exist for any fixed value of $r$.

Since the existence of trapped modes with arbitrarily large values of $q$ provides an
obstruction to the construction of a smearing function, one way around this difficulty is
to remove such modes by considering a hard momentum cutoff $\Lambda$.
Another way to think about this is that it may indeed be possible to reconstruct the
bulk data from the boundary information, but only up to a fixed momentum $\Lambda$.
As $\Lambda$ is taken larger, the reconstruction becomes more difficult, as there would
be larger amplification in going from the boundary to the bulk due to the presence of
trapped modes with larger values of $q$.  With such a cutoff, one would have good control
of the near boundary region in the bulk.  However, one would lose complete localization
in the transverse directions.

Finally, let us try to give at least a partial answer to the question
raised in the title of this paper. If we limit ourselves to a minimum
spatial resolution, local operators in the non-relativistic
CFT do indeed contain all the relevant information about fields in
the bulk of Lifshitz and other `non-relativistic' space-times. However,
full locality in the transverse direction cannot be achieved using
smearing functions only, due to the presence of modes with vanishing
boundary imprint. If and how the missing local bulk information can
be extracted from the field theory remains an interesting open question.
One possibility that comes to mind is to make use of non-local operators
in the field theory, such as Wilson-loops \cite{Susskind:1999ey}. At the very least,
our analysis demonstrates that some parts of the holographic dictionary for nonrelativistic
gauge/gravity dualities are more intricate than in the well-understood
AdS/CFT case.

\section*{Acknowledgments}
The authors would like to thank Tomas Andrade, Sheer El-Showk, Blaise Gouteraux,
Monica Guica, Peter Koroteev, Leo Pando Zayas, Ioannis Papadimitriou, Simon Ross
and Benson Way for fruitful discussions.
CAK also thanks the Cientro de Ciencias de Benasque Pedro Pascual for its hospitality
during the July 2013 Gravity Workshop.
This work was supported in part by the US Department of Energy under
grant DE-SC0007859.

\appendix

\section{\label{sec:WKB}WKB approximation }

Our
proof that smearing functions do not exist in Lifshitz and various
other nonrelativistic spacetimes relies heavily on the use of WKB
methods. While this approach in general only leads to approximate
solutions, it is nevertheless able to capture the important qualitative
behavior of the wavefunction that is needed in our analysis, up to
a finite error. It is therefore crucial to discuss this method, as
well as its limitations, in some detail.

We would like to find approximate solutions to equations of the form
\begin{equation}
\psi^{\prime\prime}+\Omega^{2}(\zeta)\psi=0,\label{eq:schroedinger wkb}
\end{equation}
with $\Omega^{2}>0$ as $\zeta\rightarrow\infty$ and $\Omega^{2}\sim-{\zeta^{-2}}$
as $\zeta\rightarrow0$. Furthermore, we shall assume that for a given
energy, there exists only one classical turning point with $\Omega^{2}\left(\zeta_{0}\right)=0$.
To capture all of these properties explicitly, we may write
\begin{equation}
\Omega^{2}=K^{2}-\frac{1}{\zeta^{2}}\left(\nu^{2}-\frac{1}{4}+\mu\left(\zeta\right)\right),
\label{eq:general omega}
\end{equation}
with $\lim_{\zeta\rightarrow0}\mu\left(\zeta\right)=0$ and $\nu>1/2$.
Notice that for $\nu\leq1/2$ the qualitative picture would
change considerably: The wavefunction becomes oscillating again close
to the boundary, which requires a different treatment. For Lifshitz
spacetime, we have $K=1$ and $\mu=\alpha\zeta^{2-k}$, where $\zeta\equiv Ex$
(see \eqref{eq:dimpot}). We now make the standard WKB-ansatz
\begin{equation}
\psi\sim\frac{1}{\sqrt{P\left(\zeta\right)}}e^{i\int d\zeta^{\prime}P\left(\zeta^{\prime}\right)}.
\end{equation}
 Plugging into \eqref{eq:schroedinger wkb}, we arrive at a differential
equation for $P\left(\zeta\right)$:
\begin{equation}
P^{2}-\Omega^{2}+\frac{1}{2}\frac{P^{\prime\prime}}{P}-\frac{3}{4}\left(\frac{P^{\prime}}{P}\right)^{2}=0.
\end{equation}
This equation can be solved perturbatively, assuming that the frequency
$\Omega^{2}$ is slowly-varying:
\begin{equation}
P^{2}=Q_{0}+\epsilon Q_{1}+\epsilon^{2}Q_{2}+\cdots,\label{eq:W expansion}
\end{equation}
where
\begin{eqnarray}
Q_{0} & \equiv & \Omega^{2},\nonumber \\
Q_{1} & \equiv & \frac{3}{4}\left(\frac{\Omega^{\prime}}{\Omega}\right)^{2}-\frac{1}{2}\frac{\Omega^{\prime\prime}}{\Omega},\label{eq: Q solutions}\\
 &\ldots,\nonumber
\end{eqnarray}
and we introduced an explicit parameter $\epsilon$ that counts the
number of derivatives and needs to be set to 1 at the end. To lowest
order, $P^{2}\approx\Omega^{2}$ and the error can be estimated by
comparing the size of the first order to the zeroth order term. Away
from the classical turning point $\zeta_{0}$, the full solution can
be written as:
\begin{equation}
\psi\left(\zeta\right)=\begin{cases}
\left(-\Omega^{2}\right)^{-\frac{1}{4}}\left[Ce^{-\int_{\zeta_{0}}^{\zeta}d\zeta^{\prime}\sqrt{-\Omega^{2}}}+De^{\int_{\zeta_{0}}^{\zeta}d\zeta^{\prime}\sqrt{-\Omega^{2}}}\right], & \,\zeta<\zeta_{0};
\\
\left(\Omega^{2}\right)^{-\frac{1}{4}}\left[ae^{i\int_{\zeta_{0}}^{\zeta}d\zeta^{\prime}\sqrt{\Omega^{2}}}+be^{-i\int_{\zeta_{0}}^{\zeta}d\zeta^{\prime}\sqrt{\Omega^{2}}}\right], & \,\zeta>\zeta_{0}.
\end{cases}\label{eq:WKB ansatz-1}
\end{equation}
As is obvious from \eqref{eq: Q solutions}, the WKB approximation
always breaks down near the turning point. As usual, this can be dealt
with by approximating the potential in the region close to $\zeta_{0}$
by a linear function
\begin{equation}
\Omega^{2}\approx\beta\left(\zeta-\zeta_{0}\right),\qquad\beta\equiv\frac{d\Omega^{2}}{d\zeta}\left(\zeta_{0}\right)>0.
\end{equation}
In this region, the solution is then given in terms of the Airy functions:
\begin{equation}
\psi_{0}\approx E_{1}\mathrm{Ai}\left(\beta^{\frac{1}{3}}\left(\zeta_{0}-\zeta\right)\right)+E_{2}\mathrm{Bi}\left(\beta^{\frac{1}{3}}\left(\zeta_{0}-\zeta\right)\right).
\end{equation}
It has the following asymptotics:
\begin{equation}
\psi_{0}\approx\begin{cases}
\frac{\left(\zeta_{0}-\zeta\right)^{-\frac{1}{4}}}{2\beta^{\frac{1}{12}}\sqrt{\pi}}\left[E_{1}e^{-\frac{2}{3}\sqrt{\beta}\left(\zeta_{0}-\zeta\right)^{\frac{3}{2}}}+2E_{2}e^{\frac{2}{3}\sqrt{\beta}\left(\zeta_{0}-\zeta\right)^{\frac{3}{2}}}\right], & \zeta\ll\zeta_{0};
\\
\frac{\left(\zeta-\zeta_{0}\right)^{-\frac{1}{4}}}{2\beta^{\frac{1}{12}}\sqrt{\pi}}\left[\left(E_{2}-iE_{1}\right)e^{i\left(\frac{\pi}{4}+\frac{2}{3}\sqrt{\beta}\left(\zeta-\zeta_{0}\right)^{\frac{3}{2}}\right)}+\left(E_{2}+iE_{1}\right)e^{-i\left(\frac{\pi}{4}+\frac{2}{3}\sqrt{\beta}\left(\zeta-\zeta_{0}\right)^{\frac{3}{2}}\right)}\right],
& \zeta\gg\zeta_{0}.
\end{cases}\label{eq:airy limits}
\end{equation}
On the other hand, for $\zeta$ close to, but not too close to $\zeta_{0}$,
the exponent in \eqref{eq:WKB ansatz-1} can be written as
\begin{equation}
\int_{\zeta_{0}}^{\zeta}d\zeta^{\prime}\sqrt{|\Omega^{2}|}\approx\begin{cases}
-\frac{2}{3}\sqrt{\beta}\left(\zeta_{0}-\zeta\right)^{\frac{3}{2}},
& \mbox{\ensuremath{\zeta}}<\zeta_{0};
\\
\frac{2}{3}\sqrt{\beta}\left(\zeta-\zeta_{0}\right)^{\frac{3}{2}},
& \zeta>\zeta_{0}.
\end{cases}\label{eq:wkb limits}
\end{equation}
 Matching \eqref{eq:airy limits} and \eqref{eq:WKB ansatz-1}, we
find
\begin{align}
C & =  \left(e^{-i\frac{\pi}{4}}a+e^{i\frac{\pi}{4}}b\right)\nonumber,
\\
D & =  \frac{i}{2}\left(e^{-i\frac{\pi}{4}}a-e^{i\frac{\pi}{4}}b\right).
\end{align}
 Near the boundary ($\zeta\ll1$), we then have
\begin{equation}
\psi(\zeta)=\frac{\zeta^{\frac{1}{2}}}{\left(\nu^{2}-\frac{1}{4}\right)^{\frac{1}{4}}}\left(Ce^{S_0(\zeta)}+De^{-S_0(\zeta)}\right),
\label{eq:psi result}
\end{equation}
 where
\begin{equation}
S_0(\zeta)\equiv\int_{\zeta}^{\zeta_{0}}d\zeta^{\prime}\sqrt{-\Omega^{2}}.
\label{eq:S integral}
\end{equation}
Hence the solution near the boundary is determined entirely in terms
of $S_0$, which is given as an integral over the effective potential.

As a check of the validity of the WKB approximation, let us determine
whether $Q_{1}$ in \eqref{eq:W expansion} remains small compared
to $Q_{0}$ for all $\zeta$. Consider the slightly more general case
where $\Omega^{2}\sim-{\zeta^{-\emm}}$ as $\zeta\rightarrow0$.
We find
\begin{equation}
Q_{1}=-\frac{\emm\left(\emm-4\right)}{16\zeta^{2}}.\label{eq:Q1 for general k}
\end{equation}
For $\emm\neq0,4$, this term blows up near the boundary. For $\emm<2$,
it blows up faster than $Q_{0}=\Omega^{2}$ itself, thus rendering
the WKB approximation invalid. For $\emm>2$, it blows up slower than
$\Omega^{2}$, so the relative error approaches zero and we should
expect WKB to yield accurate results. In the borderline case $\emm=2$,
which is the one that is interesting for us, the first order correction
is in general comparable to the zero-th order term. Hence the lowest
order approximation will a priori not give very accurate results.

Stated differently, for $\emm=2$ the perturbative expansion \eqref{eq:W expansion}
of $P$ is not consistent, since in general the order $n$ and order $n+1$
terms will mix. To avoid this mixing, we need to find a way to explicitly
move the $-1/(4\zeta^{2})$ to one lower order in the expansion.
Obviously, we could just declare
\begin{equation}
P^{2}=\Omega^{2}-\frac{1}{4\zeta^{2}}+O\left(\epsilon\right).
\end{equation}
 This is equivalent to making the somewhat ad-hoc substitution $\nu^{2}\rightarrow\nu^{2}+1/4$
in \eqref{eq:W expansion}. A more rigorous way is to perform the
following change of variables:
\begin{align}
\zeta & \equiv  e^{w},\nonumber \\
\psi & \equiv  e^{\frac{w}{2}}u.
\label{eq:new variables}
\end{align}
Then the Schr\"odinger equation reads
\begin{equation}
u^{\prime\prime}+\omega^{2}u=0,\label{eq:schroedinger new}
\end{equation}
where
\begin{equation}
\omega^{2}\equiv e^{2w}-\nu^{2}-\mu(w).
\end{equation}
It is easy to see that in these coordinates, the effective frequency
is indeed slowly varying both in the deep UV and the deep IR. In fact,
one can check that the first order term $Q_{1}$ becomes much smaller
than $\Omega^{2}$ in both limits. We see that in the new variables
\eqref{eq:new variables}, the expansion \eqref{eq:W expansion} is
consistent and the WKB solution is a good approximation everywhere,
except in the vicinity of the turning point.

Repeating the steps \eqref{eq:WKB ansatz-1}
to \eqref{eq:psi result} for \eqref{eq:schroedinger new} and changing
back to our previous variables we arrive at
\begin{equation}
\psi=\left(\frac{\zeta}{\nu}\right)^{\frac{1}{2}}\left(Ce^{S\left(\zeta\right)}+De^{-S\left(\zeta\right)}\right),
\label{eq:psi result S}
\end{equation}
 with
\begin{equation}
S\left(\zeta\right)\equiv\int_{\zeta}^{\zeta_{0}}d\zeta^{\prime}\sqrt{-\Omega^{2}+\frac{1}{4\zeta^{\prime2}}}
\,.\label{eq:S corrected}
\end{equation}
Not surprisingly, the effect of the coordinate transformation \eqref{eq:new variables}
is indeed to add an effective potential $\Delta U=1/(4\zeta^{2})$
to \eqref{eq:general omega}. Therefore, all we need to do in practice
is to replace $\nu^{2}\rightarrow\nu^{2}+1/4$. Let us emphasize
that \eqref{eq:schroedinger new} is in fact equivalent to \eqref{eq:schroedinger wkb},
so this substitution is now on a rigorous footing.

\subsection{Example: AdS ($z=1)$}

For AdS, $z=1$ and we have
\begin{equation}
\Omega^{2}=1-\frac{\nu^{2}-\frac{1}{4}}{\zeta^{2}},
\end{equation}
where
\begin{equation}
\zeta=\sqrt{E^{2}-p^{2}}\rho.
\label{eq:zetaads}
\end{equation}
Computing the integral \eqref{eq:S corrected}, we find
\begin{equation}
S\left(\zeta\right)=-\sqrt{\nu^{2}-\zeta^{2}}-\frac{\nu}{2}\log\left(\frac{\nu-\sqrt{\nu^{2}-\zeta^{2}}}{\nu+\sqrt{\nu^{2}-\zeta^{2}}}\right).
\end{equation}
 Near the boundary ($\zeta\ll\nu$),
\begin{equation}
e^{S}\approx\left(\frac{e}{2\nu}\right)^{-\nu}\zeta^{\nu}.
\end{equation}
 Plugging this result into \eqref{eq:psi result S} and rescaling back
to the original field $\phi$ we arrive at the familiar-looking result
\begin{equation}
\phi\left(x\right)=A\rho^{d-\Delta}+B\rho^{\Delta},
\end{equation}
 where $\Delta\equiv(d+1)/2+\nu$, and
\begin{align}
A & =  Ce^{-\nu}2^{\nu}\nu^{\nu-\frac{1}{2}}\left(E^{2}-p^{2}\right)^{\frac{1}{4}-\frac{\nu}{2}},\nonumber
\\
B & =  iDe^{\nu}2^{-\nu-1}\nu^{-\nu-\frac{1}{2}}\left(E^{2}-p^{2}\right)^{\frac{1}{4}+\frac{\nu}{2}}.
\end{align}
Notice that the inclusion of the correction term $\Delta U$ was crucial
to obtain the correct boundary behavior.

\subsection{Example: $z=2$ Lifshitz}

For Lifshitz with $z=2$, we have
\begin{equation}
\Omega^{2}=1-\frac{\nu^{2}-\frac{1}{4}}{\zeta^{2}}-\frac{\alpha}{\zeta}.
\end{equation}
 The classical turning point is at
\begin{equation}
\zeta_{0}=\frac{\alpha}{2}\left(1+\sqrt{1+\left(\frac{2\nu}{\alpha}\right)^{2}}\right).
\end{equation}
In this case, the WKB integral \eqref{eq:S corrected} can be evaluated to give
\begin{align}
S =&  -\sqrt{\nu^{2}+\alpha\zeta-\zeta^{2}}-\nu\log\left(\frac{\zeta\left(2\nu^{2}
+\alpha\zeta_{0}\right)}{\zeta_{0}\left(2\nu^{2}+\alpha\zeta
+2\nu\sqrt{\nu^{2}+\alpha\zeta-\zeta^{2}}\right)}\right)
\nonumber \\
&   +\frac{\pi\alpha}{4}+\frac{\alpha}{2}\mathrm{arcsin}
\left(\frac{\alpha-2\zeta}{\sqrt{4\nu^{2}+\alpha^{2}}}\right).
\end{align}
In the near-boundary limit $\zeta/\nu\rightarrow0$, with $\alpha/\nu$ held fixed, we find
\begin{equation}
e^{S}\approx\left(\frac{\sqrt{\alpha^2+(2\nu)^2}}{(2\nu)^2}\right)^{-\nu}
\exp\left[-\nu+\frac{\alpha}{2}\left(\pi-\mathrm{arctan}\left(\frac{2\nu}{\alpha}\right)\right)\right]
\cdot\zeta^{-\nu}.
\end{equation}
For $\alpha/\nu \ll 1$ this can be approximated as
\begin{equation}
e^{S}\approx\left(\frac{e}{2\nu}\right)^{-\nu}\zeta^{-\nu},
\end{equation}
which is exactly what we found in the AdS case.

Hence high energy/low
momentum modes do not ``feel'' the Lifshitz background, but instead
behave like they would in the AdS case. Those are precisely the ``free
modes'', defined in section \ref{sub:scalarslifshitz}, which only
have to tunnel through the ${\rho^{-2}}$-part of the potential.
Notice that for finite momenta, the definitions of $\zeta$ in AdS
\eqref{eq:zetaads} and Lifshitz \eqref{eq:zetadef} differ slightly.
They do however agree in the $\alpha\rightarrow0$ limit.

We are interested in the normalizable mode, which may be obtained by setting $C=0$; this
furthermore implies $D=e^{-i\frac{\pi}{4}}b$. Using \eqref{eq:psi result S}, we see that
\begin{equation}
\left.\frac{|B|}{|b|}\right|_{\mathrm{\scriptscriptstyle WKB}}=
\fft{e^\nu}{\sqrt\nu(2\nu)^{2\nu}}(\alpha^2+4\nu^2)^{\nu/2}
\exp\left[-\fft\alpha2\left(\pi-\mathrm{arctan}\left(\fft{2\nu}\alpha\right)\right)\right].
\end{equation}
This may be compared with the exact $z=2$ solution \eqref{eq:Bfrombexact}
\begin{equation}
\frac{|B|}{|b|}=2^{\fft12+\nu}\fft{|\Gamma(\fft12+\nu+i\fft\alpha2)|}{\Gamma(1+2\nu)}
e^{-\pi\alpha/4}.
\end{equation}
As an example,
we show the behavior of the WKB and exact solution as a function of $\alpha$ for $\nu=1$
in Figure~\ref{fig:nu=1z=2}.

\begin{figure}[t]
\centering
\includegraphics[width=0.6\textwidth]{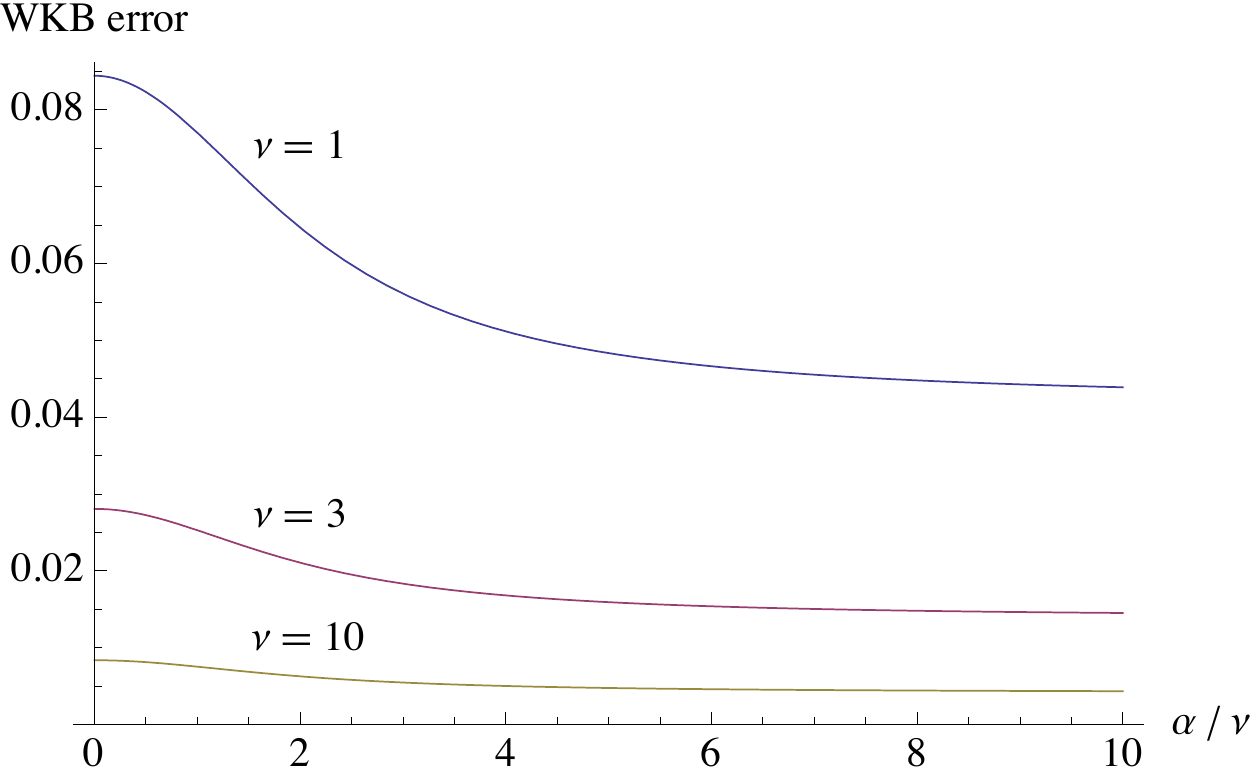}
\caption{\label{fig:z=2err}
Comparison of the WKB amplitude factor with the exact result for $z=2$ and $\nu=1$,
3 and 10.  The fractional WKB error is given by $(\eta_{\rm WKB}-\eta_{\rm exact})/\eta_{\rm exact}$,
where $\eta=|B|/|b|$.}
\end{figure}

It is straightforward to examine the behavior of the WKB and exact solutions in the small
and large $\alpha$ limits.  The $\alpha/\nu\ll1$ limit was already considered above.
In the opposite limit $\alpha/\nu\gg1$, we find instead
\begin{equation}
e^{S}\approx\left(\frac{e}{2\nu}\right)^{-2\nu}\alpha^{-\nu}e^{\frac{\alpha\pi}{2}}\zeta^{-\nu}.
\label{eq:z=2agg1}
\end{equation}
Thus we obtain
\begin{equation}
\left.\frac{|B|}{|b|}\right|_{\mathrm{{\scriptscriptstyle WKB}}}\approx\begin{cases}
\left(\frac{e}{2}\right)^{\nu}\nu^{-\left(\nu+\frac{1}{2}\right)},&\mathrm{for}\,\frac{\alpha}{\nu}\ll1;\\
\frac{e^{2\nu}}{\sqrt{\nu}\left(2\nu\right)^{2\nu}}\alpha^{\nu}e^{-\frac{\pi\alpha}{2}},&\mathrm{for}
\,\frac{\alpha}{\nu}\gg1.
\end{cases}
\end{equation}
This may be compared with the exact solution in the same limits
\begin{equation}
\frac{|B|}{|b|}\approx\begin{cases}
\frac{2^{\nu+\frac{1}{2}}\Gamma\left(\frac{1}{2}+\nu\right)}{\Gamma\left(1+2\nu\right)}, &\mathrm{for}\,\frac{\alpha}{\nu}\ll1;\\
\frac{\sqrt{4\pi}}{\Gamma\left(1+2\nu\right)}\alpha^{\nu}e^{-\frac{\pi\alpha}{2}},&\mathrm{for}\,\frac{\alpha}{\nu}\gg1.
\end{cases}
\end{equation}
This demonstrates that the WKB solution gives the correct $\alpha$ behavior for both small
and large $\alpha$.  Note that the $\nu$ dependent prefactors are different for finite $\nu$,
although they coincide in the large $\nu$ limit.  This can be seen in Figure~\ref{fig:z=2err},
where we plot the fractional difference between the WKB result and the exact solution
for several values of $\nu$.  In particular, while the asymptotic behavior $|B|/|b|\sim \alpha^\nu
e^{-\pi\alpha/2}$ is reproduced as $\alpha/\nu\to\infty$, the fractional error approaches a
constant for fixed $\nu$
\begin{equation}
\fft{\delta(|B|/|b|)}{|B|/|b|}\to
\fft{\Gamma(1+2\nu)e^{2\nu}}{\sqrt{4\pi\nu}(2\nu)^{2\nu}}-1=\fft1{24\nu}+\fft1{1152\nu^2}+\cdots.
\end{equation}
One should keep in mind, however, that this will not affect our results on the absence
of smearing functions for the Lifshitz background, as what is important is the exponential
suppression near the boundary, and not the exact form of the prefactor.

Additionally, we did not need to know the exact relationship between the coefficient $C$ in (\ref{eq:psi result S}) and the non-normalizable mode $A$; we only needed to know that setting $C=0$ forces $A=0$.  In fact, the WKB approximation cannot pick out more about the relationship between $A$ and $C$; it cannot see if there is any of the normalizable mode $B$ present in $C$ as well. If we wanted to find the Green's function, we would have trouble.  The Green's function is the response of the normalizable boundary mode to sourcing by the non-normalizable boundary mode, under infalling boundary conditions at the horizon; that is, $G= B/A|_{b=0}$.  For the exact $z=2$ solution, we find
\begin{equation}
\left.\frac{B}{A}\right|_{b=0} = (2i)^{2\nu} e^{2\pi i \nu}\frac{\Gamma(-2\nu)}{\Gamma(2\nu)}\frac{\Gamma(\frac{1}{2}+\nu+\frac{i\alpha}{2})}{\Gamma(\frac{1}{2}-\nu+\frac{i\alpha}{2})}.
\end{equation}
If we assume that the WKB term with coefficient $C$ contributes only to the non-normalizable mode with coefficient $A$, then from WKB we would find, in the large $\alpha$ limit,
\beq
\left.\frac{|B|}{|A|}\right|_{\mathrm{\scriptscriptstyle WKB}} = \left(\frac{e}{2\nu}\right)^{4\nu}\frac{\alpha^{2\nu}}{2} e^{-\alpha\pi}.
\eeq
These two expressions do not match, even when we are in a limit where the WKB error (see section (\ref{sub:Appendix error})) is small.  This mismatch, however, will not affect our analysis, because we only care about the case when the non-normalizable mode is completely turned off.

\subsection{General Lifshitz}

For the general Lifshitz case, we consider the effective potential
\begin{equation}
\Omega^2=1-\fft{\nu^2}{\zeta^2}-\fft\alpha{\zeta^k},
\end{equation}
where we recall that $k$ is related to the critical exponent by $k=2(1-1/z)$.  We restrict to the case
$z>1$, corresponding to $0<k<2$.  While the exact WKB integral may be performed numerically,
it is in fact possible to extract the asymptotic behavior in the large $\alpha$ limit.

More precisely, we note that $\Omega^2$ introduces several scales for $\zeta$, depending
on the relative importance of the three terms.  In the UV, as $\zeta\to0$, the $\nu^2/\zeta^2$
term will dominate, while in the IR, as $\zeta\to\infty$, the constant term will dominate.
If $\alpha<\nu^k$, then the $\alpha/\zeta^k$ term is not important.  In this case, the $1/\zeta^2$
piece of the potential leads to power law behavior in the UV, but no exponential suppression
in the wavefunction.  On the other hand, for $\alpha>\nu^k$, an intermediate region
$(\nu^2/\alpha)^{1/(2-k)}<\zeta<\alpha^{1/k}$ opens up, where the $\alpha/\zeta^k$ term leads
to tunneling behavior.

\begin{figure}[t]
\centering
\includegraphics[width=0.6\textwidth]{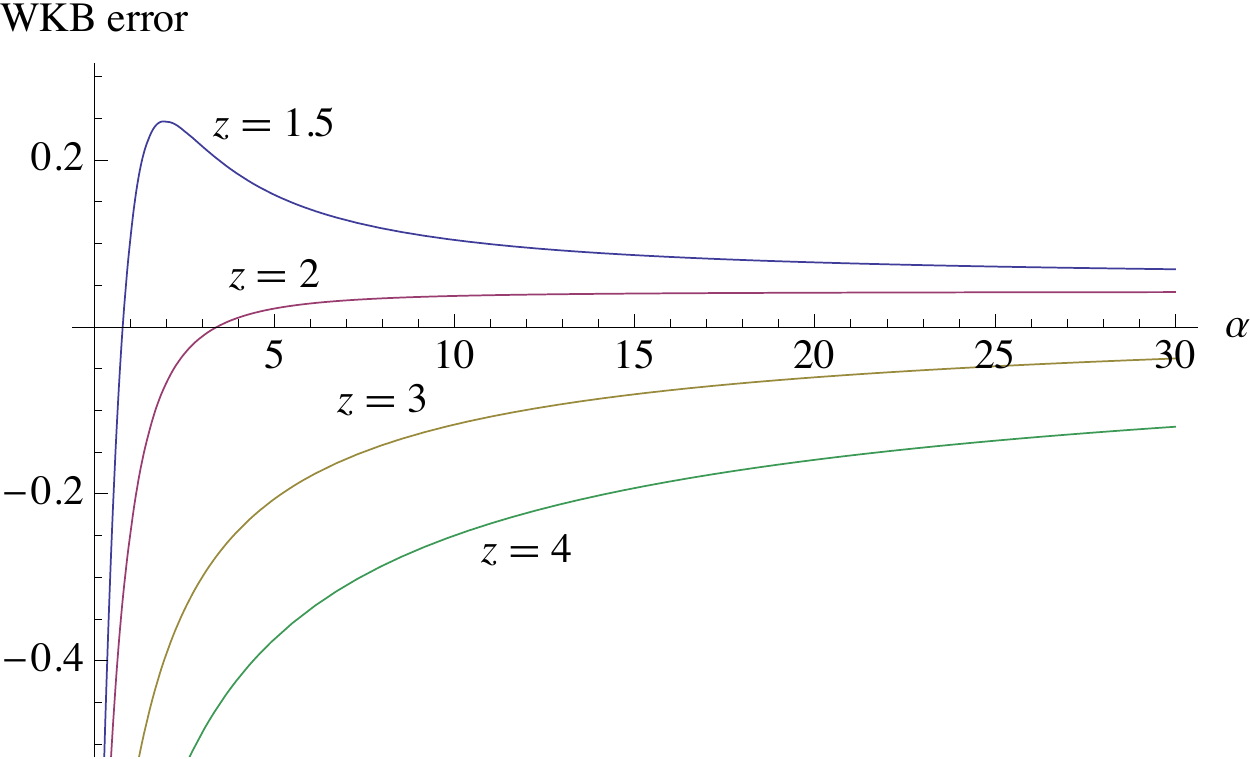}
\caption{\label{fig:lifwkb}
Comparison of the asymptotic WKB amplitude factor with the exact (numerical) result for $\nu=1$,
and $z=1.5$, 2, 3 and 4.  The fractional WKB error is given by
$(\eta_{\rm WKB}-\eta_{\rm exact})/\eta_{\rm exact}$, where $\eta=|B|/|b|$.
Note that the asymptotic WKB result \eqref{eq:WKBasymp} is only valid in the large $\alpha$ limit.
The fractional error approaches a constant (dependent on $\nu$) as $\alpha\to\infty$.}
\end{figure}

For $\alpha\gg\nu^k$, the UV and IR regions are well separated, and we may approximate the
WKB integral according to
\begin{equation}
S=\int_\zeta^{\zeta_0}d\zeta'\sqrt{\fft{\nu^2}{\zeta'^2}+\fft\alpha{\zeta'^k}-1}
\approx\int_\zeta^{\zeta_*}d\zeta'\sqrt{\fft{\nu^2}{\zeta'^2}+\fft\alpha{\zeta'^k}}
+\int_{\zeta_*}^{\zeta_0}d\zeta'\sqrt{\fft\alpha{\zeta'^k}-1}=S_1+S_2,
\end{equation}
where $(\nu^2/\alpha)^{1/(2-k)}\ll\zeta_*\ll\alpha^{1/k}$.  The first integral may be performed by
making the change of variables $u=(\alpha/\nu^2)\zeta^{2-k}$.  The result is
\begin{equation}
S_1=\fft\nu{2-k}\left.
\left[2\sqrt{1+u}+\log\fft{\sqrt{1+u}-1}{\sqrt{1+u}+1}
\right]
\phantom{\sqrt{\frac{\frac{a}{b}}{b}}}\hspace{-.8cm}
\right|_{(\alpha/\nu^2)\zeta^{2-k}}^{(\alpha/\nu^2)\zeta_*^{2-k}}.
\end{equation}
Expanding for the lower limit near zero and the upper limit near infinity gives
\begin{equation}
S_1=\fft\nu{2-k}\log\left(\fft{4\nu^2}{\alpha e^2}\right)-\nu\log\zeta
+\fft{2\sqrt\alpha}{2-k}\zeta_*^{1-k/2}\left(1-\fft{\nu^2}{2\alpha\zeta_*^{2-k}}+\cdots\right).
\end{equation}
This gives the correct near-boundary behavior
\begin{equation}
\psi_{\rm WKB}\sim\zeta^{1/2}e^{-S}\sim \zeta^{\nu+1/2}.
\end{equation}
For the second integral, we let $u=\alpha/\zeta^k$, so that
\begin{equation}
S_2=\fft{\alpha^{1/k}}k\int_1^{\alpha/\zeta_*^k}u^{-1-1/k}\sqrt{u-1}du.
\end{equation}
Although this integral can be expressed in terms of the incomplete Beta function, we only
need the expansion for large $\alpha/\zeta_*^k$.  The result is
\begin{equation}
S_2=\fft{\sqrt\pi\Gamma(1/k-1/2)}{2\Gamma(1/k)}\alpha^{1/k}-\fft{2\sqrt\alpha}{2-k}
\zeta_*^{1-k/2}\left(1-\fft{2-k}{2(2+k)}\fft{\zeta_*^k}\alpha-\cdots\right).
\end{equation}
When $S_1$ and $S_2$ are added together, the leading terms in $\zeta_*$
cancel, while the rest vanish in the asymptotic limit.  We thus obtain
\begin{equation}
\psi_{\rm WKB}\sim\sqrt{\fft\zeta\nu}e^{-S}
\sim \zeta^{\nu+1/2}\fft1{\sqrt\nu}\left(\fft{\alpha e^2}{4\nu^2}\right)^{\nu/(2-k)}
\exp\left(-\fft{\sqrt\pi\Gamma(1/k-1/2)}{2\Gamma(1/k)}\alpha^{1/k}\right).
\label{eq:WKBasymp}
\end{equation}
This agrees with \eqref{eq:z=2agg1} for $k=1$, corresponding to $z=2$.  We have
confirmed numerically that this WKB result for $\alpha\gg\nu^k$ reproduces the correct
asymptotic behavior in $\alpha$.  As an example, we show the fractional error for
several values of $z$ at fixed $\nu=1$ in Figure~\ref{fig:lifwkb}.  As in the $z=2$ case
discussed above, for fixed $\nu$, the exact prefactor is not reproduced by WKB.  However,
the exponential suppression is confirmed.

\subsection{Error analysis}\label{sub:Appendix error}

In addition to the explicit numerical analysis of the previous section, we would like to investigate the domain of validity of the WKB approximation analytically. In particular, this allows us to identify potentially problematic regions that yield a large error when integrated over, and identify when and where the WKB approximation breaks down.

In the coordinates \eqref{eq:new variables}, the effective frequency
is given by
\begin{equation}
\omega^{2}=e^{2w}-\alpha e^{\left(2-k\right)w}-\nu^{2}.\label{eq:omegasq(y)}
\end{equation}
The relative error can be estimated by
\begin{align}
\frac{Q_{1}}{Q_{0}} = &\, \frac{1}{\omega^{6}}\bigg[\frac{1}{4}e^{4w}+\nu^{2}e^{2w}+\frac{1}{16}\alpha^{2}\left(2-k\right)^{2}e^{2\left(2-k\right)w}
\nonumber \\
   & +\frac{1}{4}\alpha\left(k^{2}+k-2\right)e^{\left(4-k\right)w}-\frac{1}{4}\nu^{2}\alpha\left(2-k\right)^{2}e^{\left(2-k\right)w}\bigg].
\end{align}
Clearly, ${Q_{1}}/{Q_{0}}\rightarrow 0$ as $w\rightarrow-\infty$,
so the WKB approximation is always valid in the deep UV. The matching
procedure near the turning point is only valid if there is some finite
overlap between the matching region, where $\omega^{2}$ is approximately
linear, and the semiclassical region, where ${|Q_{1}|}/{|Q_{0}|}\ll1$.
Let us consider two separate cases:
\begin{enumerate}
\item $\alpha\ll\nu$: We can write $\omega^{2}\approx e^{2w}-\nu^{2}$.
The condition for the potential to be approximately linear is
\begin{equation}
\frac{\left(\omega^{2}\right)^{\prime\prime}(w_{0})}{\left(\omega^{2}\right)^{\prime}(w_{0})}\left(w-w_{0}\right)\ll1.
\label{eq:condition omega linear}
\end{equation}
Since the left hand side is of order $|w-w_{0}|$, the matching region
is approximately given by $e^{w}\in\left[\nu e^{-1},\nu e\right]$.
To check if there is some overlap of this interval with the semiclassical
region, let us plug the upper and lower bound into our error estimate:
\begin{equation}
\frac{|Q_{1}|}{|Q_{0}|}\approx\begin{cases}
\frac{0.08}{\nu^{2}}, & e^{w}=\nu e^{-1};
\\
\frac{0.21}{\nu^{2}}, & e^{w}=\nu e.
\end{cases}
\end{equation}
We see that for small $\nu$ (more precisely, for $\nu\lesssim{1}/{2}$),
the error becomes of order one and there is no overlap between the
matching region and the semiclassical region. In this case, the matching
procedure fails.
\item $\alpha\gg\nu$: We can write $\omega^{2}\approx e^{2w}-\alpha e^{(2-k)w}$
for $w$ near the turning point at $e^{w_{0}}\approx\alpha^{1/k}$.
The condition \eqref{eq:condition omega linear} now gives $e^{w}\in\left[\alpha^{1/k}e^{-1},\alpha^{1/k}e\right]$
and the error at the boundary points is ${Q_{1}}/{Q_{0}}\sim\alpha^{-2/k}\cdot\mathrm{const}.$
Hence for $\alpha$ large enough the matching always yields good results.
\end{enumerate}
Even though for large $\alpha$ the matching procedure works for
all $\nu$, one needs to be more careful: As we have seen previously, there are three different
regimes of $\zeta$, corresponding to each of the three terms in \eqref{eq:omegasq(y)}
dominating. In the region where $\alpha e^{\left(2-k\right)w}$ dominates,
the relative error grows as $w$ decreases (see \eqref{eq:Q1 for general k}).
If $\nu=0$, the error continues to grow to infinity as we approach
the boundary. However, for $\nu\neq0$,
the ${\nu^{2}}/{\rho^{2}}$ part of the potential takes over at
$\alpha e^{\left(2-k\right)w}\sim\nu^{2}$, and the relative error
decreases again. Hence there is a local maximum of order
\begin{equation}
\frac{|Q_{1}|}{|Q_{0}|}\approx\frac{3}{32\nu^{2}}.
\end{equation}
For small $\nu$, the WKB approximation breaks down in this region.
We speculate that since $\alpha e^{\left(2-k\right)w}\sim\nu^{2}$
is precisely where the potential changes from ${p^{2}}/{\rho}$
to ${\nu^{2}}/{\rho^{2}}$ behavior, there is some nontrivial
mixing between growing and decaying modes that the WKB approximation
cannot account for. This mixing is stronger for small $\nu$, as
the difference between the relevant exponents, $\Delta_{+}-\Delta_{-}=2z\nu$,
becomes small. Nevertheless, we can conclude that our WKB approximation can be trusted as long as $\nu\gtrsim{1}/{2}$. Most importantly, the approximation becomes more and more accurate at large $\alpha/\nu$, which is precisely the regime we are interested in.


\end{document}